\documentclass[11pt,a4paper]{article} 	

\usepackage{jheppub}
\usepackage{graphicx}
\usepackage{amssymb}
\usepackage{amsmath}
\usepackage{tikz}
\usepackage{array}
\usepackage{caption}
\usepackage{subcaption}

\usepackage{bbold}
\newcommand{\be}{\begin{equation}}
\newcommand{\ee}{\end{equation}}
\newcommand{\bea}{\begin{eqnarray}}
\newcommand{\eea}{\end{eqnarray}}
\def\bse{\begin{subequations}}
\def\ese{\end{subequations}}

\def\IZ{\relax\ifmmode\hbox{Z\kern-.4em Z}\else{Z\kern-.4em Z}\fi}

\def\del{{\partial}}

\def\presub{\vspace{.5cm} \noindent}

\def\bi{\begin{itemize}} \def\ei{\end{itemize}}

\def\({\left(} \def\){\right)}
\def\[{\left[} \def\]{\right]}
\def\<{\left<} \def\>{\right>}

\title{The vacuum seagull: evaluating a 3-loop Feynman diagram with 3 mass scales}
\author{Philipp Burda, Barak Kol and Ruth Shir \\
{\it The Racah Institute of Physics, The Hebrew University of Jerusalem,\\ Jerusalem 91904, Israel}\\
{\tt philipp.burda, barak.kol, ruth.shir@mail.huji.ac.il}
}

\abstract{We study a 3-loop 5-propagator Feynman Integral, which we call the vacuum seagull, with arbitrary masses and spacetime dimension using the Symmetries of Feynman Integrals method.  It is our first example with potential numerators. We determine the associated group $G \subset GL(3)$ which happens to be 5 dimensional and the associated set of 5 differential equations. $G$ is determined by a geometric approach which we term ``current freedom". We find the generic $G$-orbit to be co-dimension 0 and hence the method is maximally effective, and the diagram reduces to a line integral over simpler diagrams. For a reduced parameter space with 3 mass scales we are able to present explicit results in terms of special functions. This might be the first such example.}

\begin{document}
\maketitle

\section{Introduction}
Feynman diagrams and their evaluation have been known for long and much is known about them, yet the Symmetries of Feynman Integrals (SFI) method \cite{Kol:2015gsa} is a recent contribution which is rather general and natural. The SFI method considers the total parameter space of a diagram of fixed topology, composed of all possible masses and kinematical invariants of the external momenta. The parameter space is found to foliate into orbits of a continuous group $G$ which is naturally associated with the diagram. Within each leaf the Feynman integral obeys a set of linear partial differential equations, whose solution reduces to a line integral over simpler diagrams. Depending on the diagram the foliation leaves can range from being co-dimension 0, which is  maximally effective, to being dimension 0 (point-like) and hence useless. 

\begin{figure}
\begin{center}
\includegraphics[scale=0.5]{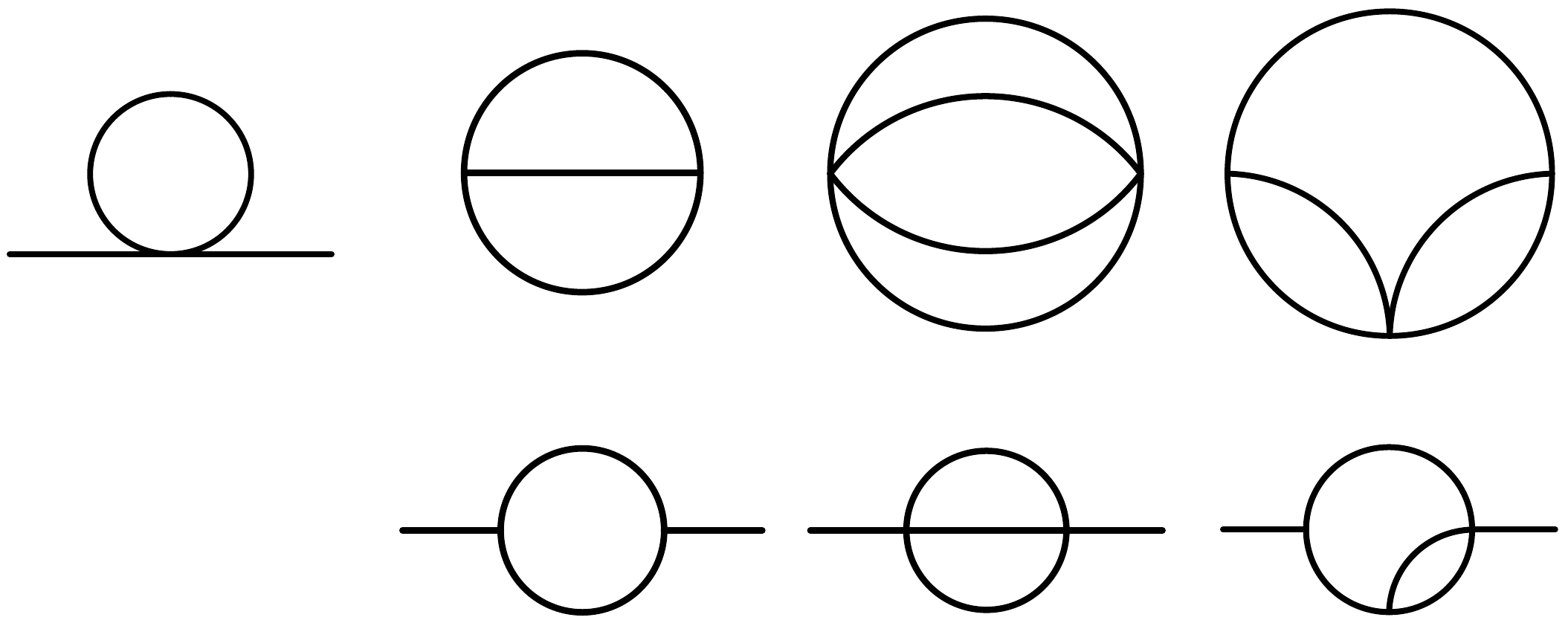}
\caption{Topologies of relevant Feynman diagrams. The upper row represents vacuum diagrams ordered by number of loops and propagators; each of these diagrams is the vacuum closure of the corresponding diagram at the bottom row.}
\label{diagram_order}
\end{center}
\end{figure}

Currently our research group is following a program to apply the SFI method to several specific diagrams in order to demonstrate it and refine it. For this purpose diagrams are naturally ordered by edge contraction since contracted diagrams appear as source terms in the SFI equation set of parent diagrams, and hence it is reasonable to proceed step by step, see Figure~\ref{diagram_order}. The 1-loop 1-propagator diagram (tadpole) is immediate. The 1-loop 2 leg  diagram (bubble) was studied in \cite{Kol:2016veg} where the orbit co-dimension was found to be zero, and a new derivation was found for the known expression 
with general parameters.
The vacuum 2-loop  diagram (diameter) is studied in \cite{diameter} wherein just as the previous case, the orbit co-dimension is zero and a new derivation is supplied for the known results which depends on all 3 masses.

\begin{figure}
\centering
\begin{subfigure}[b]{.35\textwidth}
\centering
\includegraphics[scale=0.25]{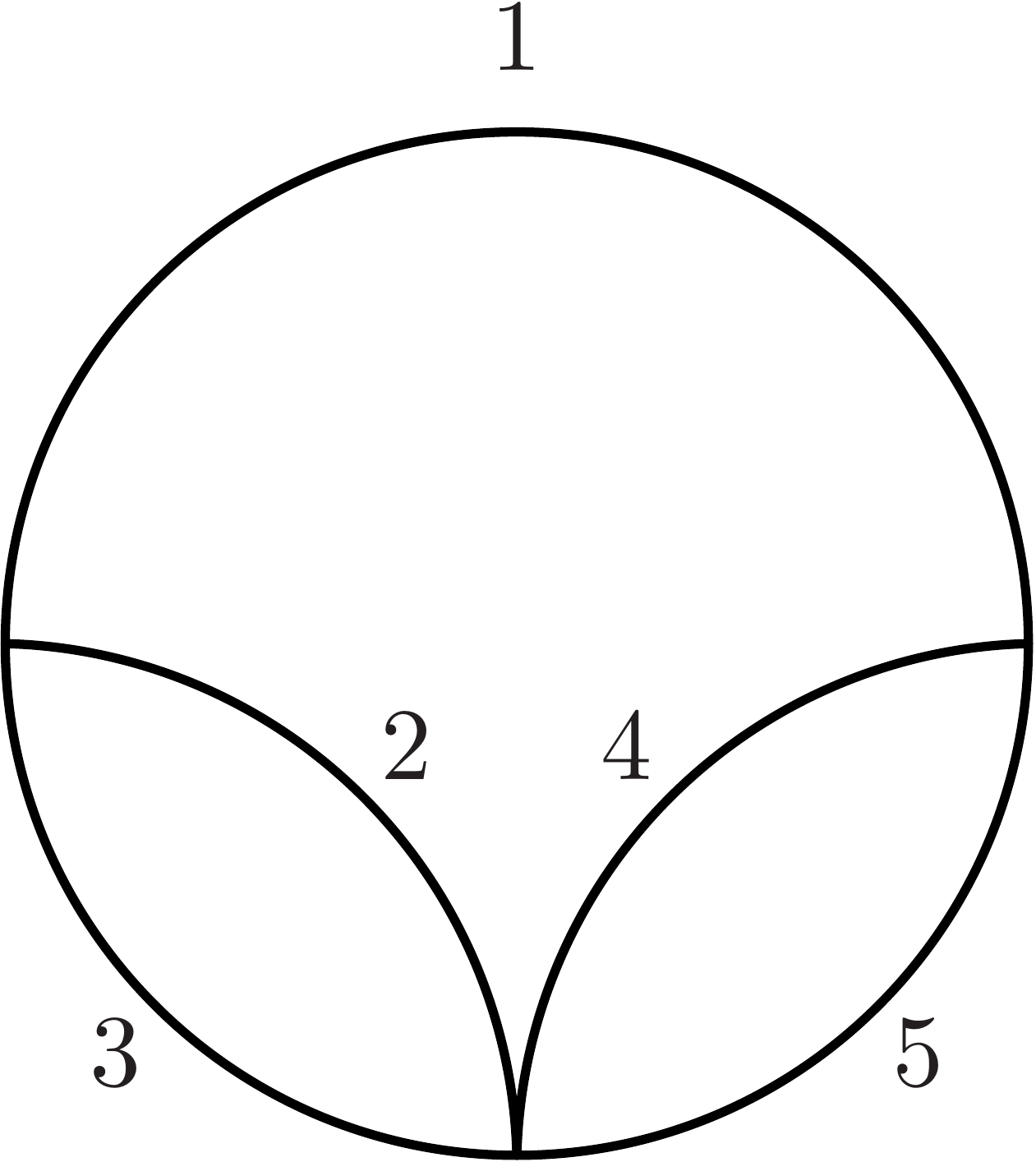}
\caption{}
\label{diagrammass}
\end{subfigure}%
\begin{subfigure}[b]{.35\textwidth}
\centering
\includegraphics[scale=0.21]{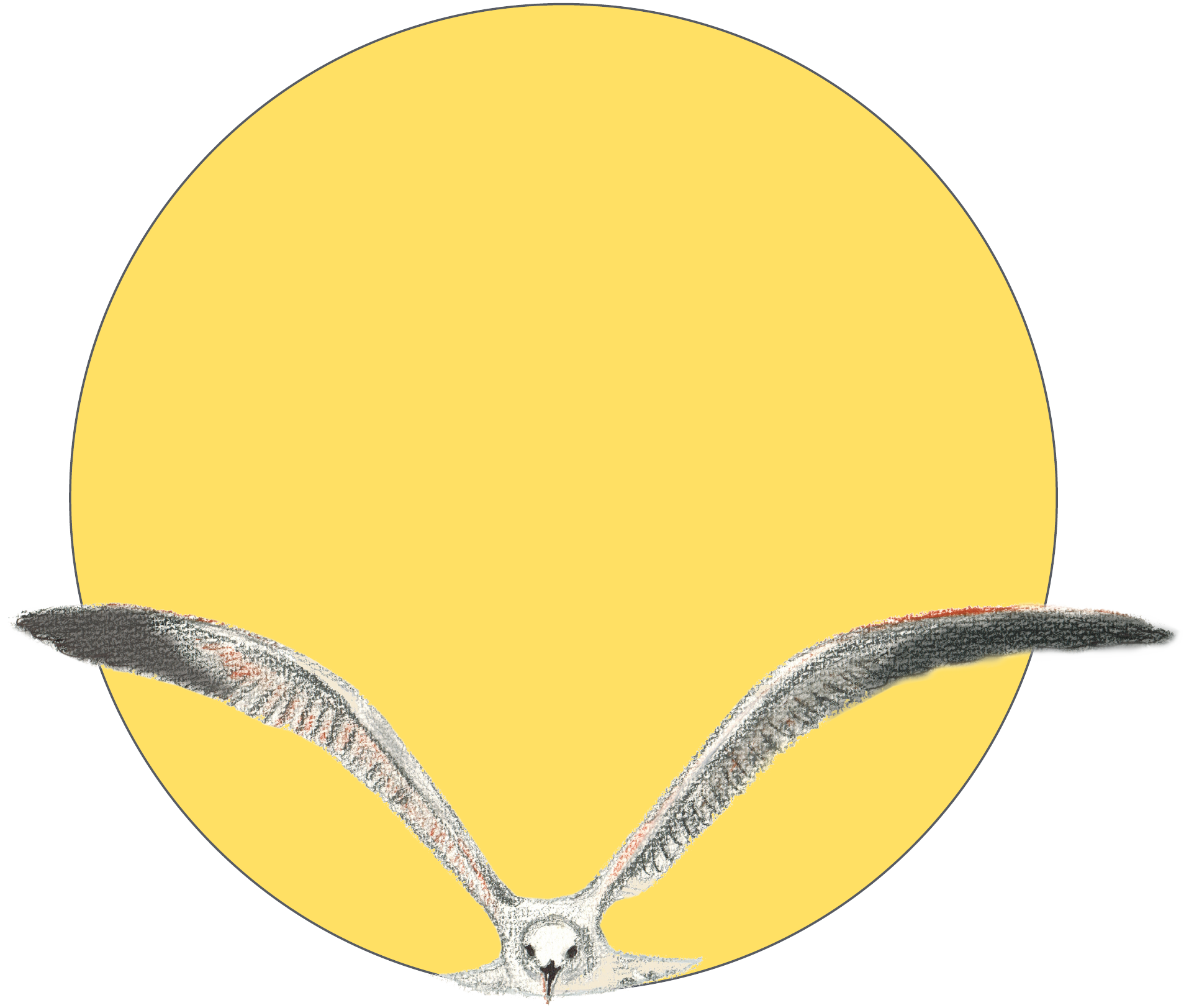}
\caption{}
\label{seagull}
\end{subfigure}%
\begin{subfigure}[b]{.35\textwidth}
\centering
\raisebox{10pt}{\includegraphics[scale=0.3]{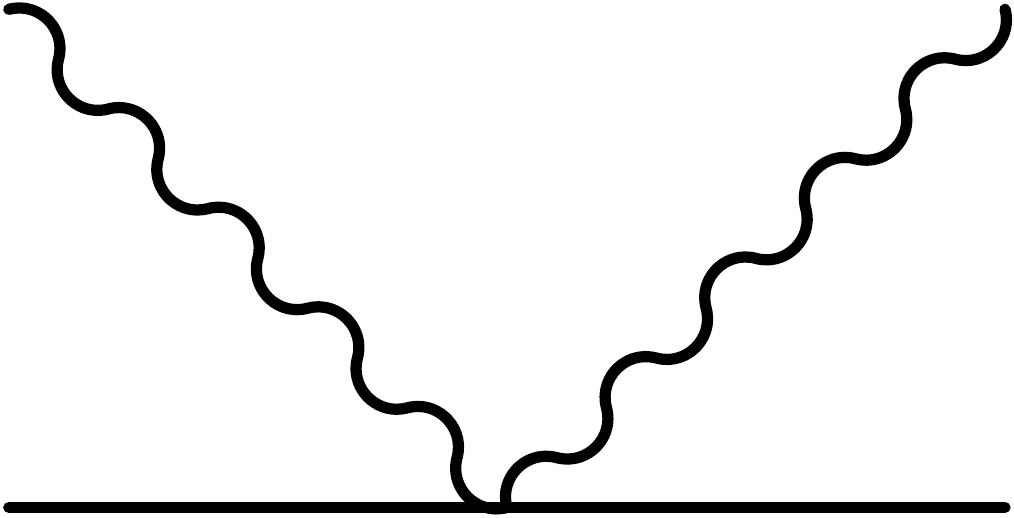}}
\caption{}
\label{seagull_vertex}
\end{subfigure}%
\caption{(a) The vacuum seagull with our labelling choice.  Illustration (b) of a seagull in front of a setting sun and (c) the standard seagull vertex explain together the motivation behind the term vacuum seagull.}
\label{vacuum_seagull}
\end{figure}

In this paper we study a 3-loop 5-propagator vacuum diagram shown in Figure~\ref{diagrammass}. We did not find a standard term for it in the literature so we felt free to refer to it as the vacuum seagull diagram, and the reason is illustrated in Figure~\ref{vacuum_seagull}. 
 Referring to the contraction order, this diagram can be contracted to two types of diagrams: contracting edge 2 (or equivalently 3, 4 or 5) factorizes into the diameter times the tadpole and hence is known, while contracting edge 1 results in a 2-vertex 4-propagator ``watermelon'' diagram which we chose to leave for future study, since the associated sunset diagram is known to involve elliptic dilogarithms (in 2d), see \cite{sunset} and references therein. 

As usual the complexity of the integral increases with the number of non-zero mass scales. 3-loop vacuum integrals with one mass scale have been solved, and in some special cases two scales are also known analytically  (see references within \cite{Freitas:2016zmy,Martin:2016bgz}). 
In particular \cite{Davydychev:2000na} considered the vacuum seagull in an $\epsilon$ expansion around $d=4$ with one mass scale; 
two arbitrary mass scales were studied in \cite{Chung:2002vp, Kalmykov:2005hb} in an $\epsilon$ expansion.  \cite{Freitas:2016zmy, Bauberger:2017nct} studied this diagram with general masses in an $\epsilon$ expansion and used dispersion relations to reduce the computation to a one dimensional integral suitable for numerical integration. 
\cite{Martin:2016bgz} studied the vacuum seagull around $d=4$ by using the method of differential equations (essentially the same as SFI) and introduced a computer package which solves an associated set of ordinary differential equations. 
 
 In the widely used Integration By Parts (IBP) method \cite{Chetyrkin:1981qh} Feynman Integrals are first reduced to master integrals (MIs). The perspective of the SFI method is somewhat different: in IBP we wish to express integrals with all possible indices as a linear combination of the MIs, while in SFI we can obtain higher indices through differentiation of the general expression with respect to parameters, namely using it as a generating function. In fact, at generic points in parameter space the vacuum seagull (having all indices equal to unity) is a master integral itself, while the other master integrals correspond to simpler diagrams (contractions of the vacuum seagull), see \cite{Martin:2016bgz}. At some special locus of points in parameter space the vacuum seagull can be expressed as a combination of simpler diagrams (see Section \ref{sec:alg}) and hence it is not an MI over there.
 
This paper is organized as follows. We start by reviewing the SFI method in Section~\ref{sec:SFI}. In Section~\ref{sec:eq} we apply the method to the vacuum seagull diagram and determine the SFI group $G$, the SFI equation set and the group structure.
The $G$-orbits are found to be co-dimension zero, and hence the SFI method reduces the dependence on all mass parameters to a line integral over simpler diagrams.  The algebraic locus is a subvariety of the parameter space where the differential SFI equation set  degenerates and becomes algebraic and hence the integral can be written as a linear combination of simpler diagrams. In Section \ref{sec:alg} we determine the algebraic locus and the solutions on it. In Section \ref{sec:lineint} we solve the SFI equation set in terms of an explicit line integral. So far the discussion applies to the most general masses. In Section \ref{sec:3d} we focus our attention to a specific 3 mass scale subspace in parameter space where the source functions (integrals associated with simpler diagrams) are known explicitly in terms of special functions. In this case we are able to present a formula in terms of a 1 dimensional integral and moreover we are able to solve it explicitly in terms of (rather rare) special functions. Section \ref{sec:checks} gives various checks for the results of Section \ref{sec:3d}. Finally in Section \ref{sec:summary} we offer a summary of our results.

\section{Symmetries of Feynman Integrals (SFI) method - general}
\label{sec:SFI}
\subsection*{Current freedom and the SFI group}
Here we shall present the geometry underlying the definition of the SFI group $G$ for a general diagram, and review the derivation of the SFI system of equations. 
The group $G$ was defined in \cite{Kol:2015gsa} and was termed ``the numerator-free sub-group''. The current perspective was originally developed in the context of the vacuum seagull and was briefly introduced in \cite{Kol:2016hak}, Section 2.

We start by setting the notation for a general diagram,  limiting ourselves to vacuum diagrams for concreteness, though the discussion generalizes also to the presence of external legs. 

Consider a general $L$ loop Feynman integral
\begin{equation}\label{generic}
I(x_1,x_2,\dots ,x_P,d)=\int \frac{d^d l_1 \dots d^dl_L}{(k_1^2-x_1)(k_2^2-x_2)\dots(k_P^2-x_P)}
\end{equation}
where $x_i=m_i^2$ are the masses on the propagators, $\{l_1,\dots,l_L\}$ are the loop momenta (or currents) and $k_i=C_{ir}l_r$ with $i=1,\dots,P$, $r=1,\dots ,L$ and $C_{ir}$ a matrix of constants, are the propagator momenta. The value of this integral is invariant under invertible linear transformations of the loop momenta $\tilde{l}_a=L_{ab}l_b$ where $L_{ab}\in GL(L)$ and the SFI group $G$ is a particular subgroup of $GL(L)$ as we will see below.

Given a choice of loop currents $l_r, r=1,\dots,L$ the space of currents is given by the span \be
 C := Sp \{ l_r \}_{r=1}^L~.
 \label{def:C}
 \ee

Next we recall the definition of the space of \emph{potential numerators} $M$ given by \be
M := Q/ S
\label{def:num}
\ee
where \be
 Q = S^2(C) \equiv Sp \{ l_r \cdot l_s \}_{r,s=1}^{L}
\ee
is the space of all quadratic Lorentz scalars made out of currents, and \be
S = Sp \{ k_i^2 \}_{i=1}^P
\ee
is the space of all squares, spanned by the squares of edge (or propagator) currents $k_i, ~i=1,\dots, P$. Since each edge current $k_i$ is a linear combination of loop currents, we can identify \be
S \subset Q ~.
\ee

We define the SFI group $G$ as the subgroup of $GL(L)$ which preserves $S$ within $Q$ as a subspace, but not necessarily pointwise.

\subsection*{SFI system of equations}
Now we proceed to the review of the derivation of the SFI system of equations by the current freedom perspective, which consists of viewing the the SFI equations as being generated by infinitesimal loop currents redefinitions of the form 
\be
 l_r \to l_r+\epsilon \, l_s \label{trans}
\ee
for arbitrary $r,s=1,\dots,L$. This is a linear redefinition of the basis for loop currents, and as such is interpreted as the freedom to choose such a basis. For this reason we refer to the above mentioned geometric method as ``current freedom". This redefinition is the basis for both the IBP method \cite{Chetyrkin:1981qh} and the method of differential equations (DE) \cite{Kotikov1a,Kotikov1b,Kotikov1c,Remiddi1997,GehrmannRemiddi1999}. Hence we consider SFI to be a refinement of these methods. The SFI method will be applied to the vacuum seagull diagram in the next section.

The current redefinition (\ref{trans}) operates on the integration variables and hence leaves the integral invariant as discussed above. Any such variation translates into a recursive equation in the usual IBP set-up.
However we limit ourselves to linear transformations in the subgroup $G$. This guarantees that no Lorentz scalars which are not in $S$ will be generated and hence allows to recast the recursion relation into a differential equation in $x$ space (an SFI equation). The transformation (\ref{trans}) induces the following transformation on the Feynman integral (\ref{generic})
\begin{equation}
\delta I(x_1,\dots , x_p) =\int  d^d l_1\dots d^d l_L \big(l_s \frac{\partial}{\partial l_r} +d\delta_{rs}\big)\frac{1}{(k_1^2-x_1)(k_2^2-x_2)\dots(k_P^2-x_P)}~.
\end{equation}
On the other hand the variation vanishes being a change of integration variables. This can be written as the following equation
\begin{equation}\label{geqs}
\int d^d l_1 \dots d^dl_L \frac{\partial}{\partial l_r}l_s \frac{1}{(k_1^2-x_1)(k_2^2-x_2)\dots(k_P^2-x_P)}=0~.
\end{equation}
Considering only the linear transformations (\ref{trans}) which are in $G$ we find the relevant operations $\frac{\partial}{\partial l_r}l_s$ to act on the Feynman integral. For each of these operators we get a PDE in parameter space from (\ref{geqs}) as we will describe now. The left hand side of (\ref{geqs}) gives a sum of terms: the first comes from operating by $\frac{\partial}{\partial l_r}$ on $l_s$ and gives back the original integral multiplied by $d$ when $r=s$; the other terms come from operating on the integrand by $\frac{\partial}{\partial l_r}$ and gives a sum of terms. In each of these terms one propagator from the edges of the loop $r$ is squared and a numerator of the form $l_s \frac{\partial k_i(l_r)}{\partial l_r}$ is generated. Since we have considered only transformations from $G$ we can express the numerator in terms of propagators and linear combinations of $x$'s. A squared propagator $i$ in the denominator is interpreted as taking a derivative by $x_i$, that is, $\frac{1}{(k_i^2-x_i)^2}=\frac{\partial}{\partial x_i} \frac{1}{(k_i^2-x_i)}$. In this way we obtain three types of terms from equation (\ref{geqs}): the first are constants multiplying the original integral $I(x)$, the second are linear combinations of $x$ derivatives of $I(x)$ with linear combinations of $x$ as coefficients, the third are degenerate integrals where one propagator is eliminated and another is squared; we call these integrals ``sources" and denote them by $J(x)$. By doing this for each operator in $G$ we get a system of PDEs of the form
\begin{equation}
c_a\,I(x)-M_{ai}(x)\frac{\partial}{\partial x_i}I(x)=J_a(x)
\end{equation}
where $a$ enumerates the equations and $i=1,\dots ,P$. The matrix $M(x)$ contains the linear $x$ dependent coefficients of $\frac{\partial I}{\partial x}$ and $J(x)$ denotes sums of degenerate integrals. 

We believe that the SFI equation set is essentially not different from DE. SFI added value includes insisting on considering the whole set, exposing the underlying geometry (identification of $G$ and the foliation of parameter space) and finally steps towards the solution of the set (including the reduction to the line integral).
In the next section we find the system of PDEs for a specific three loop vacuum diagram.

\section{SFI group and system of equations for vacuum seagull}
\label{sec:eq}

In this section we shall apply the SFI method to the vacuum seagull diagram to obtain the SFI group $G$ and the SFI equation set.

Consider the three-loop five-propagator diagram shown in Figure~\ref{diagrammass}. 
For general masses on the propagator legs this diagram is given by the following integral 
\begin{eqnarray}\label{vs}
I(x_1,\dots,x_5,d)&=&
\int \frac{d^d l_1 d^d l_2 d^d l_3}{(l_1^2-x_1)(l_2^2-x_2)((l_1+l_2)^2-x_3)(l_3^2-x_4)((l_1+l_3)^2-x_5)}
\label{original_integral}
\end{eqnarray}
where $x_i \equiv m_i^2$. The value of the integral depends only on the parameter space 
\be
X:=\{x_i\}_{i=1}^5
\ee
and on the spacetime dimension $d$. 

\subsection*{SFI group for vacuum seagull}
Let us apply the geometrical procedure from the previous section. The vacuum seagull has~3 loops ($L=3$) and hence $\rm{dim}(Q) = 6$ while the number of propagators is 5, or equivalently $\rm{dim}(S) = 5$. Therefore $\rm{dim}(M) = \rm{dim}(Q)-\rm{dim}(S)=1$. In all previously studies cases (the diameter and the bubbled diagrams) there were no potential numerators (namely  $\rm{dim}(M) = 0$) and so this is our first example with (non-trivial) potential numerators. 

We shall see here how the current freedom perspective can facilitate the determination of the group $G$ for the diagram under study. According to the conventions of (\ref{original_integral}) the square subspace is given by \be
 S = Sp \{ l_1^2,\, l_2^2,\, l_3^2, \(l_1+l_2\)^2,\, \(l_1 + l_3\)^2\, \}~.
 \label{def:Svs}
 \ee
Since $S$ is a 5-dimensional subspace of a 6d space, it is more convenient to characterize it by its perpendicular subspace $S_\perp$, which is a subspace of $Q^*$, the dual of $Q$. More concretely, the space dual the space of loop currents $C$ is spanned by $\del^r\equiv \del/\del l_r, ~r=1,2,3$, and accordingly \be
 Q^* = Sp \{ \del^r \cdot \del^s \}_{r,s=1}^3~.
\ee 
Since $s_\perp := \del^2 \del^3$ annihilates all generators of $S$ (\ref{def:Svs}) one sees that \be
S_\perp := Sp \{ \del^2 \del^3 \}~.
\ee
Transformations which preserve the 1d subspace $S_\perp$ must rescale $s_\perp$, and hence $G$ can be decomposed as \be
G = \{ r\,  \mathbb{1} \} \times G_1 
\ee
where the first factor is proportional to the unit matrix $\mathbb{1}$ and $G_1$ is defined to preserve $s_\perp$. Since $s_\perp$ is a quadratic form of signature $(+-0)$, $G_1$ can be 
identified with a Lorentz group of a degenerate 3d space (with one timelike direction, one spacelike and one null). Concretely $G_1$ is seen to consist of transformations of the form 
\be
 \[\begin{array}{ccc} \del^1 & \del^2 & \del^3 \end{array} \]
 = 
  \[\begin{array}{ccc} \del'^1 & \del'^2 & \del'^3 \end{array} \]
 \[ \begin{array}{ccc} 
a 	& 0	 & 0 \\
*	& e^w & 0 \\
*	&	0 &  e^{-w} 
\end{array} \]
\ee
where $a \neq 0$ so that the transformation is invertible and $*$ denotes an arbitrary entry, independent of all other $*$s.
After allowing back the unit matrix we find that $G$ is generated by the transformations 
\be
 \[\begin{array}{ccc} \del^1 & \del^2 & \del^3 \end{array} \]
 = 
  \[\begin{array}{ccc} \del'^1 & \del'^2 & \del'^3 \end{array} \]
 \[ \begin{array}{ccc} 
* 	& 0	 & 0 \\
* 	& *	 & 0 \\
*	& 0	&  * 
\end{array} \]
\label{trans1}
\ee
(as long as the diagonal entries are non-zero).

This action translates back to generators in $C$ space as 
\be
\delta \[\begin{array}{c} l_1 \\  l_2 \\ l_3 \end{array} \]
 = 
   \[ \begin{array}{ccc} 
* 	& 0	 & 0 \\
* 	& *	 & 0 \\
*	& 0	 &  * 
\end{array} \]
\[\begin{array}{c} l_1 \\  l_2  \\ l_3 \end{array} \]
\label{generators1}
\ee
where now the $*$'s are completely unconstrained. This can be seen as follows. A linear transformation of the form $\del^r=\del'^s M_s^{~r}$ implies that $M_s^{~r} = \del l'_s / \del l_r$ and hence $l_r = (M^{-1})_r^{~s}\, l'_s$. Given a linear transformation of form (\ref{trans1}) both its inverse and the associated generators have a similar form leading to (\ref{generators1}).

The form of the generators given by (\ref{generators1}) means that $G$ has 5 generators, namely $\rm{dim}(G)=5$, and they can be listed as follows $\frac{\del}{\del l_1} l_1,\, \frac{\del}{\del l_2} l_2,\, \frac{\del}{\del l_2} l_1,\, \frac{\del}{\del l_3} l_3,\, \frac{\del}{\del l_3} l_1$. Note that while both $\rm{dim}(G)$ and the number of propagators happen to be 5 for the vacuum seagull, this property is a coincidence and does not generalize to other diagrams.

\subsection*{SFI system of equations for vacuum seagull}
Applying these $5$ generators to (\ref{original_integral}) results in 5 linear PDEs which are of the form
\begin{subequations} \label{PDEs}
\begin{equation}
c_a I(x) - M_{ai}(x)\, \frac{\partial}{\partial x_i} I(x) = J_a(x)
\end{equation}
where $a=1,\dots,5$ enumerates the equations and $i=1,\dots,5$ is the propagator index. $c_a$ are constants that can depend on $d$; $M(x)$ is a $5\times 5$ matrix with entries linear in $x$ and $J_a(x)$ are combinations of integrals which originate from $I(x)$ by eliminating one propagator and squaring another. Their explicit forms are given by
\begin{equation}\label{c}
c = \begin{pmatrix}
d-4 \\ d-3 \\ 0 \\ d-3 \\ 0
\end{pmatrix},
\quad
J(x)=\begin{pmatrix}
J_{13}-J_{23}+ J_{15}-J_{45} \\
-J_{13}+J_{23} \\
-J_{12} +J_{32}+J_{13}-J_{23}\\
-J_{15}+J_{45}\\
-J_{14}+J_{54}+J_{15}-J_{45}
\end{pmatrix}
\end{equation}
\begin{equation} \label{M}
M(x)=\begin{pmatrix}
2x_1 & 0 & x_1-x_2+x_3 & 0 &  x_1-x_4+x_5 \\ 
0 & 2x_2 & -x_1+x_2+x_3  & 0 & 0 \\
0  & -x_1-x_2+x_3 & x_1-x_2+x_3 & 0 & 0 \\
0 & 0 & 0 & 2x_4 & -x_1+x_4+x_5 \\
0 & 0 & 0 & -x_1-x_4+x_5 & x_1-x_4+x_5
\end{pmatrix}
\end{equation}
\end{subequations}
where $J_{ij}$ are the integrals obtained from Equation (\ref{original_integral}) by contracting the propagator $i$ and squaring the propagator $j$; their explicit form is given in (\ref{j12})-(\ref{j54}).

The topologies of the $J$ integrals are given by the two possible degenerations of $I(x)$:
\begin{equation*}
\mathrm{Degen}\left[ \raisebox{-12pt}{\includegraphics[scale=0.2]{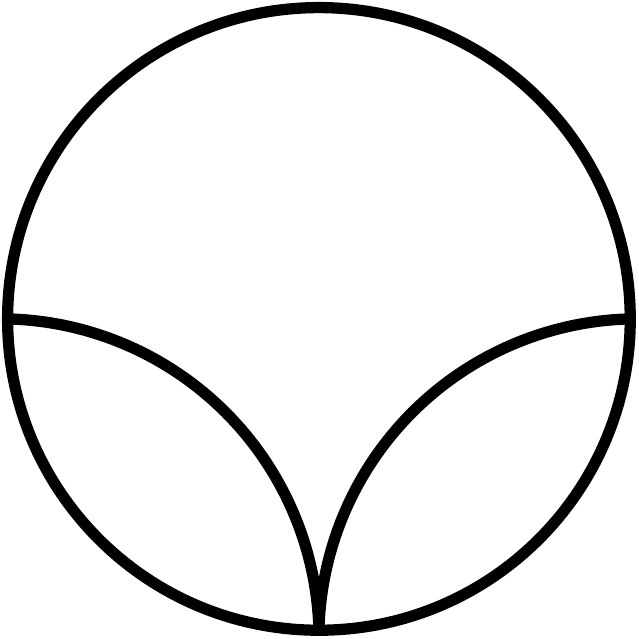}}\right]=\left\{\raisebox{-12pt}{\includegraphics[scale=0.2]{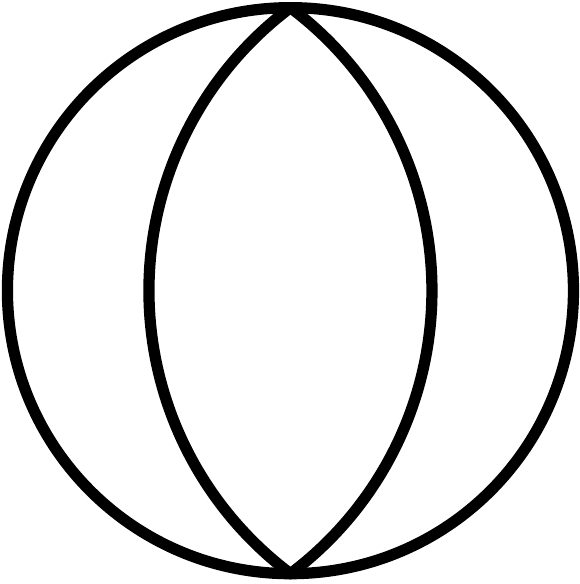}}\, , \; \raisebox{-7pt}{\includegraphics[scale=0.2]{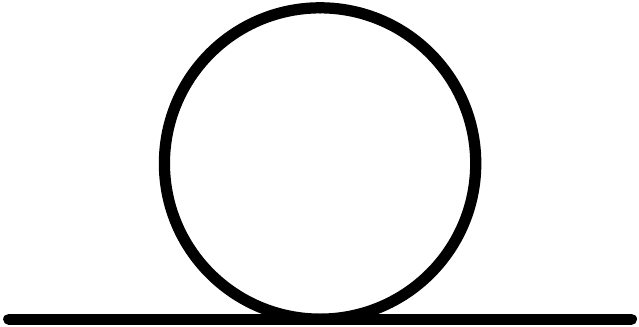}} \times \raisebox{-9pt}{\includegraphics[scale=0.2]{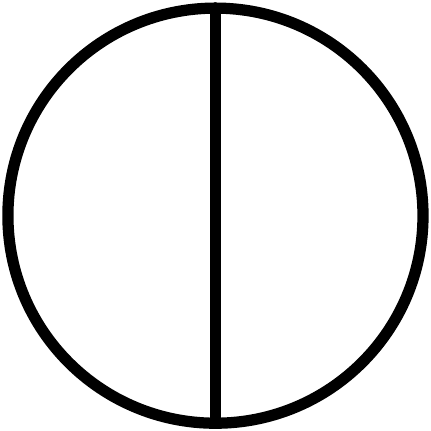}} \right\}.
\end{equation*}

\subsection*{Group orbits within parameter space}
The set of differential equations (\ref{PDEs}) defines the differential operators \begin{equation}
\label{do}
T_a(x) \equiv M_{ai}(x)\frac{\partial}{\partial x_i}, \quad a=1,\dots,5
\end{equation}
which form a representation of $G$ on the parameter space $X$. 

Computing the determinant $\det[M(x)]$ (see (\ref{detm}) below) we find that at a generic point in the 5 dimensional $X$ space the 5 operators $T_a$ are linearly independent, and hence the generic orbit is 5 dimensional, or in other words
\begin{equation} \label{codim}
\mathrm{codim}(G\,\mathrm{orbits})=0~.
\end{equation}
This statement tells us that it is enough to know the value of the integral at a specific point (or at most at a discrete set) and then perform a line integral to any desired point in $X$ space. In the next section we will discuss special hypersurfaces which we call ``the algebraic locus" on which the set $\{T_a\}$ is degenerate and where we can find algebraic solutions for $I(x)$ in terms of $J_a(x)$.

The $T_a$ operators (\ref{do}) inherit the group structure of the fundamental set of operators 
$\frac{\del}{\del l_1} l_1,\, \frac{\del}{\del l_2} l_2,\, \frac{\del}{\del l_2} l_1,\, \frac{\del}{\del l_3} l_3,\, \frac{\del}{\del l_3} l_1$.
In particular we find that they satisfy the commutation relations
\begin{equation}
\label{comrel}
[T_3,T_2]=T_3, \quad [T_5,T_4]=T_5~.
\end{equation}
The operator $T_1+T_2+T_4$ is responsible for transformations in the radial direction in $X$ space (the dimension operator). Using this observation and the commutation relations (\ref{comrel}) we find that our group decomposes as $G=U(1)_{\mathrm{radial}} \times U(1)^2 \rtimes U(1)^2$.

\section{Algebraic locus}
\label{sec:alg}
Before proceeding to the solution of the SFI system of equations (\ref{PDEs}), it is worth discussing a special set of hypersurfaces within the parameter space on which the system is degenerate and the value of the integral is determined algebraically. 

\begin{figure}
\begin{center}
\begin{tikzpicture}[scale=3]
\draw (0,0) -- (1/2,0.866025) -- (1,0) -- (0,0);
\draw [blue] (0.5,0.288675) circle [radius=0.288675];
\draw [blue] (0,0) -- (1,0);
\node [below left] at (0,0) {$x_2$};
\node [above] at (1/2,0.866025) {$x_1$};
\node [below right] at (1,0) {$x_3$};

\node at (1.25,0.433013) {$\times$};

\draw (1.5,0) -- (2,0.866025) -- (2.5,0) -- (1.5,0);
\draw [blue] (2,0.288675) circle [radius=0.288675];
\draw [blue] (1.5,0) -- (2.5,0);
\node [below left] at (1.5,0) {$x_4$};
\node [above] at (2,0.866025) {$x_1$};
\node [below right] at (2.5,0) {$x_5$};
\end{tikzpicture}
\caption{Schematic projective representation of $X$ space: the circle in the left triangle is $\lambda(x_1,x_2,x_3)=0$, the circle in the right triangle is $\lambda(x_1,x_4,x_5)=0$ and $x_1=0$ is represented by $(x_2 x_3)\cup(x_4 x_5)$.}
\label{parameter_space}
\label{default}
\end{center}
\end{figure}
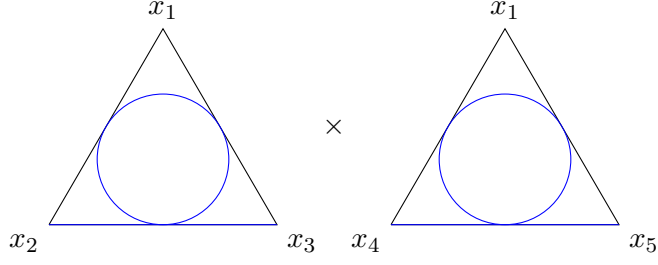

For this diagram the matrix $M_{ai}(x)$ defined in (\ref{M}) is square and therefore we can compute its determinant
\begin{equation}\label{detm}
\det[M(x)]=-2x_1 \lambda(x_1,x_2,x_3) \lambda(x_1,x_4,x_5)
\end{equation}
where
\begin{equation}
\lambda(x,y,z)=x^2+y^2+z^2 -2 x y-2 x z-2 y z
\end{equation}
is the Heron or K\"all\'en invariant. Since $\det[M(x)] \neq 0$ generically, we see that generic $G$ orbits are 5 dimensional.
Furthermore, we see that the set $\{T_a\}$ is degenerate on the following hypersurfaces in $X$ space: 
\begin{eqnarray}
x_1&=&0 \label{x1=0}\\
\lambda_{123} \equiv \lambda(x_1,x_2,x_3) &=& 0 \label{l123=0} \\
\lambda_{145} \equiv \lambda(x_1,x_4,x_5) &=& 0 \label{l145=0}~.
\end{eqnarray}
As we shall explain immediately below, (\ref{PDEs}) reduces to algebraic equations for $I(x)$ on these hypersurfaces, therefore we will refer to them as ``algebraic locus" hypersurfaces. They are represented schematically in Figure~\ref{parameter_space}.

Let us multiply (\ref{PDEs}) by the matrix of polynomials $\tilde{M}(x):=\text{det}[M(x)]M^{-1}(x)$
\begin{equation}
\tilde{M}(x)_{ia}c_a I-\text{det}[M(x)]\partial_i I=\tilde{M}(x)_{ia}J_a~.
\end{equation}
Setting $\text{det}[M(x)]=0$ we arrive at a system of algebraic equations for $I(x)$:
\begin{eqnarray}\label{alg_loc}
&&(d-3)\begin{pmatrix}
x_{145}x_{154}\lambda_{123}+ x_{123}x_{132}\lambda_{145}-\frac{d-4}{d-3}\lambda_{123}\lambda_{145}\\ 
-2x_1 x_{132}\lambda_{145} \\ -2x_1 x_{123}\lambda_{145} \\ -2x_1 x_{154}\lambda_{123} \\ -2x_1 x_{145}\lambda_{123}
\end{pmatrix}I \\ \nonumber
&&\quad=\begin{pmatrix}
\lambda_{123}[2x_5 x_{145}(J_{45}-J_{15})+2x_4x_{154}(J_{54}-J_{14})]+(4,5)\leftrightarrow(2,3)\\-2x_1 x_{231}\lambda_{145}(J_{12}-J_{32})+4x_1 x_3\lambda_{145}(J_{13}-J_{23})\\-2x_1 x_{321}\lambda_{145}(J_{13}-J_{23})+4x_1 x_2\lambda_{145}(J_{12}-J_{32})\\-2x_1x_{451}\lambda_{123}(J_{14}-J_{54})+4x_1 x_5\lambda_{123}(J_{15}-J_{45})\\
-2x_1 x_{541}\lambda_{123}(J_{15}-J_{45})+4x_1x_4\lambda_{123}(J_{14}-J_{54})
\end{pmatrix}
\end{eqnarray}
where we have defined $x_{ijk}:=x_i-x_j-x_k$.
Now we can analyze the solutions of these equations on each branch of the algebraic locus (\ref{x1=0}-\ref{l145=0}) separately.

\presub {\bf Solutions on $x_1=0$}.
To find the algebraic solution on $x_1=0$ we simply set $x_1=0$ in (\ref{alg_loc}) to get
\begin{eqnarray} 
(d-2)I(0,x_2,x_3,x_4,x_5) &=&
\frac{2x_2 \big(J_{32}-J_{12}\big)- 2x_3 \big(J_{23}-J_{13}\big)}{x_2-x_3} \nonumber \\
& & + \frac{2x_4 \big(J_{54}-J_{14}\big)-2x_5\big(J_{45}-J_{15}\big)}{x_4-x_5} \label{Ix1}~.
\end{eqnarray}

From here we can examine several simpler cases. For example, when additionally $x_3=x_5=0$ we get 
\begin{equation}
(d-2)I(0,x_2,0,x_4,0) = 2(J_{32}-J_{12}) + 2(J_{54}-J_{14})\label{I135} ~.
\end{equation}
We confirm that this result coincides with the answer obtained via \texttt{FIRE} \cite{Smirnov:2014hma}.

\presub {\bf Solutions on $\lambda_{123}=0$ and $\lambda_{145}=0$}.
On $\lambda_{145}=0$ we can find the solution for $I(x)$ by setting $\lambda_{145}=0$ in (\ref{alg_loc}).  Note that adding to the solution a multiple of $\lambda_{145}$ does not change the value of $I(x)$ and therefore we can get different but equivalent expressions for $I(x)$ on the algebraic locus hypersurfaces.
The nicest form for the expression we got is
\begin{equation} \label{I145}
(d-3)I(x)=\sqrt{\frac{x_4}{x_1}}(J_{54}-J_{14})+\sqrt{\frac{x_5}{x_1}}(J_{45}-J_{15})~.
\end{equation}
The solution on $\lambda_{123}=0$ can be read from (\ref{I145}) by exchanging $(4,5)\leftrightarrow (2,3)$. We also have compared this solution with the answer of \texttt{FIRE} for the point $x_1=x_4=m^2$, $x_3=x_5=0$ and $x_2=M^2$ and found full agreement.

\presub {\bf An alternative combined expression on $\det[M(x)]=0$}.
From (\ref{alg_loc}) a general solution for $I(x)$ on the algebraic locus can be obtained
\begin{equation}
I(x)=-\frac{2\big[\big( x_1(x_1-x_3)+x_2(x_2-x_3)\big)(J_{12}-J_{32})+(2 \leftrightarrow 3)\big]\lambda_{145}+(2,3)\leftrightarrow (4,5)}{(d-3)(3x_1^2+(x_2-x_3)^2)\lambda_{145}+(2,3)\leftrightarrow (4,5)-(d-4)\lambda_{123}\lambda_{145}}~.
\end{equation}
From this general expression one can read off the solutions (\ref{Ix1}) and (\ref{I145}) as well as solutions on intersections of the algebraic locus hypersurfaces.

\section{Solution}
\label{sec:lineint}
It is always possible to reduce the solution to the SFI equation set into a line integral \cite{Kol:2015gsa}. Here we apply this general procedure to (\ref{PDEs}). In the next section we will perform the line integral for a 3 dimensional parameter space. 

The homogeneous solution of (\ref{PDEs}) is
\be
\label{homo-sol}
I_0(x)=x_1^{1-\frac{d}{2}}\left[\lambda(x_1,x_2,x_3)\lambda(x_1,x_4,x_5)\right]^{\frac{d-3}{2}}~.
\ee
The complete solution could be presented in the following form
\be \label{compsol}
I(x)=I_0(x)c(x)
\ee
where $c(x)$ should satisfy
\be
\begin{pmatrix}\del_{x_1}  \\ \del_{x_2}  \\ \del_{x_3}  \\ \del_{x_4}  \\ \del_{x_5}  \end{pmatrix}c(x)=\frac{1}{I_0(x)}
\begin{pmatrix}\frac{\lambda_{123}\lambda_{145}J_1-\{(x_1-x_2+x_3)\lambda_{145}\big((x_1+x_2-x_3)J_2+2x_2J_3\big)+(2,3)\leftrightarrow(4,5)\}}{2x_1\lambda_{123}\lambda_{145}}  \\ \frac{(x_1-x_2)(J_2+J_3)+x_3(J_2-J_3)}{\lambda_{123}}\\\frac{2x_2J_3+(x_1+x_2-x_3)J_2}{\lambda_{123}} \\ \frac{(x_1-x_4)(J_4+J_5)+x_5(J_4-J_5)}{\lambda_{145}}\\ \frac{2x_4J_5+(x_1+x_4-x_5)J_4}{\lambda_{145}}  \end{pmatrix}~.
\ee
The value of $c(x)$ at any point $B$ in the parameter space can be calculated by a line integral from a chosen starting point $A$ to $B$ along an arbitrary path $\gamma$
\bea
c(B)&=&\int_\gamma \Big[\frac{\lambda_{123}\lambda_{145}J_1-\{(x_1-x_2+x_3)\lambda_{145}\big((x_1+x_2-x_3)J_2+2x_2J_3\big)+(2,3)\leftrightarrow(4,5)\}}{2x_1\lambda_{123}\lambda_{145}I_0} dx_1\nonumber \\
& &+ \frac{(x_1-x_2)(J_2+J_3)+x_3(J_2-J_3)}{\lambda_{123}I_0} dx_2+\frac{2x_2J_3+(x_1+x_2-x_3)J_2}{\lambda_{123}I_0}dx_3 \nonumber \\
& &+ \frac{(x_1-x_4)(J_4+J_5)+x_5(J_4-J_5)}{\lambda_{145}I_0} dx_4 +\frac{2x_4J_5+(x_1+x_4-x_5)J_4}{\lambda_{145}I_0} dx_5 \Big] +c(A) \label{cb}~.
\eea
From (\ref{compsol}) it follows that
\be
c(A)=\frac{I(A)}{I_0(A)} \label{ca}
\ee
it is therefore useful to choose the starting point $A$ where $I(A)$ is easy to calculate by other methods (for example where all but one of the masses are zero).
Given (\ref{cb}) the complete solution at any point $B$ is given by 
\be \label{gensolb}
I(B)=I_0(B)c(B)~.
\ee
Now we shall present a concrete form for $I(B)$ by choosing a starting point $A$ and a path $\gamma$ from $A$ to a general point $B$ in $X$ space. We choose $A$ to be the point where $x_1=m^2$ and all the rest of the masses are zero, and $\gamma$ to be a straight line with constant $x_1$ such that
\begin{subequations} \label{trajectory}
\bea
A&=&m^2(1,0,0,0,0) \\
B&=&m^2(1,x,u,y,v)  
\eea
and it is parameterized by
\be
(x_1(t),x_2(t),x_3(t),x_4(t),x_5(t))=m^2(1,x\,t,u\,t,y\,t,v\,t), \quad 0\leq t \leq 1 ~.
\ee
\end{subequations}
Along this path $dx_1=0$ and the first term in (\ref{cb}) vanishes. We can therefore write
\bea \label{mainlineint}
I(B)&=&\big(\lambda(1,x,u)\lambda(1,y,v)\big)^{\frac{d-3}{2}} \nonumber\\
& \times & \Bigg\{\int_0^1 \left[\lambda(1,xt,ut) \lambda(1,yt,vt)\right]^{\frac{3-d}{2}}\nonumber \\
& &\Big[\frac{x(1-xt)(-J_{12}+J_{32})+u(1-ut)(-J_{13}+J_{23})+xut(-J_{12}+J_{32}-J_{13}+J_{23})}{\lambda(1,xt,ut)} \nonumber\\
& &+ \frac{y(1-yt)(-J_{14}+J_{54})+v(1-vt)(-J_{15}+J_{45})+yvt(-J_{14}+J_{54}-J_{15}+J_{45})}{\lambda(1,yt,vt)} \Big]dt\nonumber\\
& & +\frac{I(m^2,0,0,0,0)}{(m^2)^{3d/2-5}}\Bigg\} \label{lineint}~.
\eea
Note that the 2 terms in the integrand are related by symmetry, and that the $m^2$ dependence comes from the sources.
 In the next section we compute this integral on a reduced parameter space where $x_3=x_5=0$.

\subsection{Solution with 3 different mass scales}
\label{sec:3d}
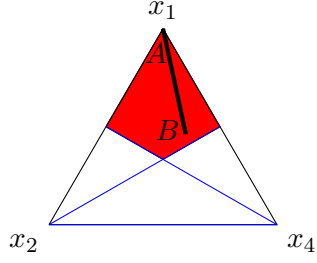
\begin{figure}
\begin{center}
\begin{tikzpicture}[scale=3]
\draw [fill=red] (0.25,0.433013) -- (1/2,0.866025) -- (0.75,0.433013) -- (1/2, 0.288675) -- (0.25,0.433013);
\draw (0,0) -- (1/2,0.866025) -- (1,0) -- (0,0);
\draw [blue] (0,0) -- (0.75,0.433013);
\draw [blue] (1,0) -- (0.25,0.433013);
\draw [blue] (0,0) -- (1,0);
\draw [ultra thick] (1/2,0.866025) -- (0.6,0.4);
\node [below left] at (0,0) {$x_2$};
\node [above] at (1/2,0.866025) {$x_1$};
\node [below right] at (1,0) {$x_4$};
\node [ ] at (0.47 ,0.76) {$A$};
\node [ ] at (0.52,0.42) {$B$};
\end{tikzpicture}
\caption{Projective representation of the 3 dimensional parameter space $(x_1,x_2,x_4)$, namely the subspace $x_1+x_2+x_4=m^2$. The blue lines represent the algebraic locus surfaces. The thick line represents the trajectory from $A=m^2(1,0,0)$ to $B=m^2(1,x,y)$ with parameterization $(x_1(t),x_2(t),x_4(t))=m^2(1,x\,t,y\,t)$ and $0 \leq t \leq 1$. The red shaded kite represents the region where the solution (\ref{vs_3m}) is manifestly non-singular.}
\label{traj}
\label{default}
\end{center}
\end{figure}

On the subspace with $x_3=x_5=0$ all of the sources $J_{ij}$ are known in terms of special functions and we can perform the 1-dimensional integral (\ref{lineint}) to get $I(x_1,x_2,0,x_4,0)$ in terms of special functions as we will explain below. This 3 dimensional subspace $(x_1,x_2,x_4)$ is depicted in Figure~\ref{traj} where the blue lines represent the algebraic locus surfaces $x_1=0$, $x_1=x_2$ and $x_1=x_4$ and the thick black line represents the path (\ref{trajectory}) with $x_3=x_5=0$.

Setting $x_3=x_5=0$ (equivalently $u=v=0$) into (\ref{mainlineint}) we get the following integral
\bea \label{lineint3}
I(m^2,m^2x,0,m^2y,0)&=&\left[(1-x)(1-y)\right]^{d-3}\Big\{\int_0^1 \Big[\frac{x(-J_{12}+J_{32})}{(1-xt)^{d-2}(1-yt)^{d-3}} \nonumber\\
& &+ \frac{y(-J_{14}+J_{54})}{(1-yt)^{d-2}(1-xt)^{d-3}}\Big]dt +\frac{I(m^2,0,0,0,0)}{(m^2)^{3d/2-5}}\Big\}\nonumber ~.
\eea
The expressions for $J_{12}$, $J_{14}$, $J_{32}$ and $J_{54}$ with $x_3=x_5=0$ are given by (\ref{redj12}-\ref{redj54}). On the chosen trajectory (\ref{trajectory}) they have the form
\begin{eqnarray}
J_{12}(t)&=&-e^{-\frac{3}{2} (d-4)\gamma_E}(m^2)^{\frac{3d}{2}-5} (y\,t)^{\frac{3d}{2}-5} \frac{\big(\Gamma(d/2-1)\big)^2}{\Gamma(d/2)} \nonumber \\
&\times&\Big[ \Gamma \big(5-\frac{3d}{2}\big)\Gamma(4-d)\Gamma \big(\frac{d}{2}-2\big) {_{2}F_{1}}\big(5-\frac{3d}{2},4-d,3-\frac{d}{2} \big| \frac{x}{y}\big)\nonumber \\
&+& \Big(\frac{x}{y}\Big)^{d/2-2}\Gamma(3-d)\big(\Gamma(2-d/2)\big)^2{_{2}F_{1}}\big(3-d,2-\frac{d}{2},\frac{d}{2}-1 \big| \frac{x}{y}\big)\Big] \label{expj12}
\end{eqnarray}

\begin{eqnarray}
J_{32}(t)&=&-e^{-\frac{3}{2} (d-4)\gamma_E}(m^2)^{\frac{3d}{2}-5} (x\,t)^{\frac{d}{2}-2}\frac{\Gamma(2-d/2)}{\Gamma(d/2)}\nonumber \\
&\times&\Big[ \Gamma(3-d)\Gamma\big(2-\frac{d}{2}\big)\big(\Gamma\big(\frac{d}{2}-1\big)\big)^2(1-y\,t)^{d-3}\nonumber \\
&+&(y\,t)^{\frac{d}{2}-1}\Gamma\big(1-\frac{d}{2}\big)\Gamma\big(2-\frac{d}{2}\big)\Gamma\big(\frac{d}{2}-1\big){_{2}F_{1}}\big(1,2-\frac{d}{2},\frac{d}{2} \big| y\,t\big)\Big]~. \label{expj32}
\end{eqnarray}
$J_{14}(t)$ and $J_{54}(t)$ are obtained from (\ref{expj12}) and (\ref{expj32}) respectively by $x \leftrightarrow y$.
At the starting point $A$ the value of the integral is easily calculated to be
\be \label{inta}
I(m^2,0,0,0,0)= -  e^{-\frac{3}{2} (d-4)\gamma_E} (m^2)^{\frac{3d}{2}-5}\frac{\Gamma(5-3d/2)(\Gamma(2-d/2))^2(\Gamma(d/2-1))^4\Gamma(3d/2-4)}{(\Gamma(d-2))^2\Gamma(d/2)}~.
\ee

Plugging the explicit expressions (\ref{expj12}), (\ref{expj32}) into (\ref{lineint3}) we organize the result according to the different $t$ integrals to be performed:
\begin{subequations}\label{explicit}
\begin{eqnarray}
I(m^2,m^2x,0,m^2y,0)&=&(m^2)^{\frac{3d}{2}-5}\big((1-x)(1-y)\big)^{d-3} \nonumber \\
& & \Big\{ c_1(x,y,d) \int_0^1 \frac{t^{\frac{3d}{2}-5}}{(1-x\,t)^{d-2}(1-y\,t)^{d-3}}dt \nonumber \\
& & +c_2(x,y,d) \int_0^1\frac{t^{\frac{d}{2}-2}}{(1-x\,t)^{d-2}}dt \nonumber \\
& & +c_3(x,y,d) \int_0^1 \frac{t^{d-3}{_{2}F_{1}}\big(1,2-\frac{d}{2},\frac{d}{2} \big| y\,t\big)}{(1-x\,t)^{d-2}(1-y\,t)^{d-3}}dt \nonumber \\
& &+ x \leftrightarrow y + c_4(d) \Big\} \label{explicit_int}
\end{eqnarray}
where
\begin{eqnarray}
c_1(x,y,d)&=& e^{-\frac{3}{2} (d-4)\gamma_E} \frac{\big(\Gamma(d/2-1)\big)^2}{\Gamma(d/2)} \Big[ \Gamma \big(5-\frac{3d}{2}\big)\Gamma(4-d)\Gamma \big(\frac{d}{2}-2\big) {_{2}F_{1}}\big(5-\frac{3d}{2},4-d,3-\frac{d}{2} \big| \frac{x}{y}\big)\nonumber \\
& &+ \Big(\frac{x}{y}\Big)^{d/2-2}\Gamma(3-d)\big(\Gamma(2-d/2)\big)^2{_{2}F_{1}}\big(3-d,2-\frac{d}{2},\frac{d}{2}-1 \big| \frac{x}{y}\big)\Big] x\,y^{\frac{3d}{2}-5} \\
c_2(x,y,d)&=& -e^{-\frac{3}{2} (d-4)\gamma_E}\frac{\Gamma(3-d) \big(\Gamma\big(2-\frac{d}{2}\big) \Gamma\big(\frac{d}{2}-1\big)\big)^2}{\Gamma(d/2)}\, x^{\frac{d}{2}-1} \\
c_3(x,y,d)&=& -e^{-\frac{3}{2} (d-4)\gamma_E} \frac{\Gamma\big(1-\frac{d}{2}\big)\big(\Gamma\big(2-\frac{d}{2}\big)\big)^2\Gamma\big(\frac{d}{2}-1\big)}{\Gamma(d/2)} (x\,y)^{\frac{d}{2}-1} \\
c_4(d) &=& -e^{-\frac{3}{2} (d-4)\gamma_E}\frac{\Gamma(5-3d/2)(\Gamma(2-d/2))^2(\Gamma(d/2-1))^4\Gamma(3d/2-4)}{\Gamma(d/2)(\Gamma(d-2))^2} ~.
\end{eqnarray}
\end{subequations}

The first two integrals in (\ref{explicit_int}) are recognized as the Appell $F_1$ function and the hypergeometric ${_{2}F_{1}}$
\begin{eqnarray}
\int_0^1 \frac{t^{\frac{3d}{2}-5}}{(1-x\,t)^{d-2}(1-y\,t)^{d-3}}dt &=&  \frac{2}{3d-8}F_1(3d/2-4,d-2,d-3,3d/2-3 | \, x,y) \\
\int_0^1\frac{t^{\frac{d}{2}-2}}{(1-x\,t)^{d-2}}dt &=& \frac{2}{d-2} {_{2}F_{1}}(d/2-1,d-2,d/2 \big|\,x)~.
\end{eqnarray}
The third integral in (\ref{explicit_int}) we find to be the Kamp\'e de F\'eriet function defined in~(\ref{Kampedeferiet})
\begin{eqnarray}
& &\int_0^1 \frac{t^{d-3}{_{2}F_{1}}\big(1,2-\frac{d}{2},\frac{d}{2} \big| y\,t\big)}{(1-x\,t)^{d-2}(1-y\,t)^{d-3}}dt = \int_0^1 \frac{t^{d-3}{_{2}F_{1}}\big(\frac{d}{2}-1,d-2,\frac{d}{2} \big| y\,t\big)}{(1-x\,t)^{d-2}}dt \nonumber \\
& &= \frac{1}{d-2} \sum_{k=0}^\infty \sum_{n=0}^\infty  \frac{(d-2)_{k+n} (d-2)_k (d/2-1)_n (d-2)_n}{(d-1)_{k+n}(d/2)_n}\frac{x^k}{k!}\frac{y^n}{n!} \nonumber \\
& & = \frac{1}{d-2} F^{\substack{1:1;2\\ 1:0;1}}\left( \begin{matrix}d-2:d-2;d/2-1,d-2 \\ d-1: \quad ;d/2\end{matrix} \Big| x,y\right) ~.
\end{eqnarray}

We thus obtain an explicit expression for the vacuum seagull diagram on the reduced parameter space with 3 arbitrary masses:
\begin{eqnarray}\label{vs_3m}
I(m^2,m^2x,0,m^2y,0)&=& e^{-\frac{3}{2} (d-4)\gamma_E}(m^2)^{\frac{3d}{2}-5}\big((1-x)(1-y)\big)^{d-3} \nonumber \\
& & \Big\{ c_1(x,y,d) \frac{2}{3d-8}F_1(3d/2-4,d-2,d-3,3d/2-3 | \, x,y)\nonumber \\
& & +c_2(x,y,d) \frac{2}{d-2} {_{2}F_{1}}(d/2-1,d-2,d/2 |\,x)\nonumber \\
& & +c_3(x,y,d)  \frac{1}{d-2} F^{\substack{1:1;2\\ 1:0;1}}\left( \begin{matrix}d-2:d-2;d/2-1,d-2 \\ d-1: \quad ;d/2\end{matrix} \Big| x,y\right)  \nonumber \\
& &+ x \leftrightarrow y + c_4(d) \Big\}~.
\end{eqnarray}
where all symbols were defined above. This expression is manifestly non-singular when $0< x,y < 1$, namely when the end point $B$ is within the red shaded region in Figure \ref{traj}.

\subsection{Checks}
\label{sec:checks}
\subsection*{Comparison for $x_2=0$}
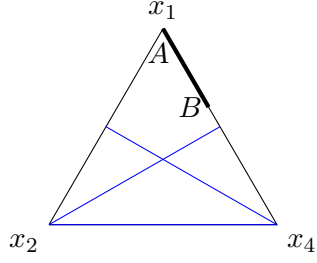
\begin{figure}
\begin{center}
\begin{tikzpicture}[scale=3]
\draw (0,0) -- (1/2,0.866025) -- (1,0) -- (0,0);
\draw [blue] (0,0) -- (0.75,0.433013);
\draw [blue] (1,0) -- (0.25,0.433013);
\draw [blue] (0,0) -- (1,0);
\draw [ultra thick] (1/2,0.866025) -- (0.7,0.519615);
\node [below left] at (0,0) {$x_2$};
\node [above] at (1/2,0.866025) {$x_1$};
\node [below right] at (1,0) {$x_4$};
\node [ ] at (0.48 ,0.76) {$A$};
\node [ ] at (0.62,0.52) {$B$};
\end{tikzpicture}
\caption{In the 3 dimensional parameter space $(x_1,x_2,x_4)$ the thick line represents the trajectory from $A=m^2(1,0,0)$ to $B=m^2(1,0,y)$ with the parameterization $(x_1(t),x_2(t),x_4(t))=m^2(1,0,y\,t)$ and $0 \leq t \leq 1$.}
\label{trajx20}
\label{default}
\end{center}
\end{figure}

Setting $x_2=0$ (equivalently $x=0$) in our reduced line integral (\ref{lineint3}), we obtain the following integral to compute
\bea \label{lineint2}
I(m^2,0,0,m^2y,0)&=&(1-y)^{d-3}\Big\{\int_0^1 \frac{y(-J_{14}+J_{54})}{(1-yt)^{d-2}(1-xt)^{d-3}} dt +\frac{I(m^2,0,0,0,0)}{(m^2)^{3d/2-5}}\Big\}~.\nonumber\\
\eea
The trajectory (\ref{trajectory}) for $x_2=0$ on the 3 dimensional subspace $(x_1,x_2,x_4)$ is shown in Figure~\ref{trajx20}.

In this case ($x_2=x_3=x_5=0$) the sources $J_{14}$ and $J_{54}$ can be calculated directly from their integral definitions (\ref{j14}) and (\ref{j54}). On the trajectory given by (\ref{trajectory}) they have the following form
\begin{eqnarray}
J_{14}(t)&=&- e^{-\frac{3}{2} (d-4)\gamma_E}(m^2)^{\frac{3d}{2}-5}
\frac{\Gamma(5 - 3 d/2) \Gamma(3 - d) (\Gamma( d/2-1))^3}{\Gamma(d/2)} (y\, t)^{3d/2-5} \label{red2j14}\\
J_{54}(t) &=&- e^{-\frac{3}{2} (d-4)\gamma_E}(m^2)^{\frac{3d}{2}-5} \frac{(\Gamma(2-d/2))^2(\Gamma(d/2-1))^2\Gamma(3-d)}{\Gamma(d/2)}(y\,t)^{d/2-2} \label{red2j54}~.
\end{eqnarray}
Plugging (\ref{red2j14}) and (\ref{red2j54}) into (\ref{lineint2}) we find
\begin{eqnarray}
I(m^2,0,0,m^2y,0)&=&- e^{-\frac{3}{2} (d-4)\gamma_E}(m^2)^{\frac{3d}{2}-5}(1-y)^{d-3}\nonumber\\
& & \times \Big[  -\frac{\Gamma(5 - 3 d/2) \Gamma(3 - d) (\Gamma( d/2-1))^3}{\Gamma(d/2)}\int_0^1 \frac{ y(y\, t)^{3d/2-5}}{(1-y\,t)^{d-2}}dt \nonumber\\
& & +\frac{(\Gamma(2-d/2))^2(\Gamma(d/2-1))^2\Gamma(3-d)}{\Gamma(d/2)}\int_0^1 \frac{ y(y\,t)^{d/2-2}}{(1-y\,t)^{d-2}}dt \nonumber\\
& &+\frac{\Gamma(5-3d/2)(\Gamma(2-d/2))^2(\Gamma(d/2-1))^4\Gamma(3d/2-4)}{(\Gamma(d-2))^2\Gamma(d/2)}\Big]
\end{eqnarray}
which we easily integrate to
\begin{eqnarray}
I(m^2,0,0,m^2 y,0)&=&-e^{-\frac{3}{2} (d-4)\gamma_E}(m^2)^{\frac{3d}{2}-5}( 1-y )^{d-3} \frac{(\Gamma( d/2-1))^2}{\Gamma(d/2)}  \nonumber \\
& & \times \Big[ - \Gamma(3 - d)\Gamma(5 - 3 d/2)  \Gamma( d/2-1)\frac{2}{3d-8}y^{3d/2-4}  {_{2}F_{1}}\big(d-2,\frac{3d}{2}-4,\frac{3d}{2}-3 \big| y\big) \nonumber \\
& & +(\Gamma(3 - d)\Gamma(2-d/2))^2\frac{2}{d-2}y^{d/2-1}{_{2}F_{1}}\big(\frac{d}{2}-1,d-2,\frac{d}{2} \big| y) \nonumber \\
& & +\frac{\Gamma(5-3d/2)(\Gamma(2-d/2))^2(\Gamma(d/2-1))^2\Gamma(3d/2-4)}{(\Gamma(d-2))^2}\Big]~. \label{vs_2m}
\end{eqnarray}
We have compared this result with direct calculation via Schwinger parameters and found complete agreement. We have also performed the line integral starting at $(x_1,x_2,x_4)=m^2(0,0,1)$. In addition we checked that in the limit $x\to0$ (\ref{vs_3m}) reduces to (\ref{vs_2m}). We note that this integral has been evaluated in closed form in \cite{Kalmykov:2011ks}, Eqs.\ (114, 116).

\subsection*{Epsilon expansion}
In this subsection we present the results for the $\epsilon$-expansion of the vacuum seagull diagram around critical dimension $d = 4$. This could serve as another check for our line integral answer (\ref{lineint3}), as we will compare our results obtained from this line integral with the results of two recent papers, \cite{Freitas:2016zmy} and \cite{Martin:2016bgz}, and will find perfect agreement.

Within the SFI approach one just needs to know the $\epsilon$-expansion of the sources, calculate the $\epsilon$-expansion of the homogeneous solution, combine them together under the line integral and perform this last integration. The general solution for 3 mass scales (\ref{lineint3}) can be presented as follows
\begin{equation}
\label{main-red}
I(B) = \int_0^1 \frac{I_0(B)(J_{12}-J_{32})x}{(1-xt)^{d-2}(1-yt)^{d-3}} dt + \left( x \leftrightarrow y \right) +\frac{I_0(B)}{I_0(A)} I(A)~.
\end{equation}
We will use this expression to study the $\epsilon$-expansion. In general, the $\epsilon$-expansion for the $I(B)$ could be written as follows\footnote{the order of a leading divergence follows from the expansion of the sources.}
\begin{equation}
I(B) = I_3(B) \epsilon^{-3} + I_2(B) \epsilon^{-2} + I_1(B) \epsilon^{-1} + I_0(B)~.
\end{equation}
Here we will discuss the first two leading divergent parts.

\presub{\bf Order $\epsilon^{-3}$}.
Using explicit expressions for the sources  (\ref{expj12}), (\ref{expj32}), one finds
\begin{eqnarray}
\frac{I_0(B)(J_{12}-J_{32})x}{(1-xt)^{d-2}(1-yt)^{d-3}} & = & \frac{m^2 x (1-x) (1-y) (3 + t y)}{6 (1- t  x)^2 (1- t y)} \epsilon^{-3} + \mathcal{O}(\epsilon^{-2}) \nonumber \\
\frac{I_0(B)}{I_0(A)} I(A) & = & \frac{1}{3} m^2 (1-x) (1-y) \epsilon^{-3} + \mathcal{O}(\epsilon^{-2}) \nonumber
\end{eqnarray}
performing the integral over $t$ from $0$ to $1$ and combining all terms together we arrive to the following very simple expression
\begin{equation}
I_3 (B) = \frac{1}{6} m^2 \left(2 + x +y \right) = \frac{1}{6} \left( 2 m_1^2 + m_2^2 +m_4^2 \right) \label{e-3}
\end{equation}
which coincide with the expressions from \cite{Freitas:2016zmy} and \cite{Martin:2016bgz}, after proper redefinitions and in the limit of $m_3 = m_5 = 0$.

\presub{\bf Order $\epsilon^{-2}$}.
In this case the expressions are a bit longer and there is no point to present them explicitly. Naively, dilogarithms appear at this order after integration. However, by using the following identity
\begin{equation}
\rm{Li}_2(z) + \rm{Li}_2(1-z) = \frac{\pi^2}{6} - \log{(z)}\log{(1-z)}
\end{equation}
we can get rid of them and after all other possible simplifications we arrive to the following answer for the $\epsilon^{-2}$ order term
\begin{eqnarray}
I_2 (B) & = & \frac{m^2}{6} \left[ 10 + 6x + 6y + 3(2+x+y) \log{m^2} + 3 x \log{x} + 3 y \log{y} \right] = \nonumber\\
& = & m_1^2 \left(\frac{5}{3} - \log{m_1^2} \right) + m_2^2 \left( 1 - \frac{1}{2} \log{m_2^2} \right) + m_4^2 \left( 1 - \frac{1}{2} \log{m_4^2} \right) \label{e-2}
\end{eqnarray}
which is again in an agreement with \cite{Freitas:2016zmy} and \cite{Martin:2016bgz}.

\subsection*{Numerical results}
The integral (\ref{lineint3}) with (\ref{expj12}) and (\ref{expj32}) can be integrated numerically for non-integer $d$. In Table~\ref{numerical} we present numerical results for various values of $d$ with $x_1=1$ and $x_2,x_4 \in [0.01,0.9]$. These values of $x_1,x_2,x_4$ cover the region colored in red in Figure~\ref{traj}. We did not cross the algebraic locus surfaces since the integral (\ref{lineint3}) diverges there and must be regularized. We have compared the numerical integral values with the values obtained from our explicit analytic solution (\ref{vs_3m}) and found complete agreement.

\begin{table}
\caption{Numerical results for various $d$. We normalized the results by setting $x_1=1$ and $x_2,x_4 \in [0.01,0.9]$. The horizontal and vertical axis represent increasing $x_2$ and $x_4$ respectively.}
\begin{center}
\begin{tabular}{c c }
$d=3.2$ & $d=3.7$ \\
\includegraphics[scale=0.4]{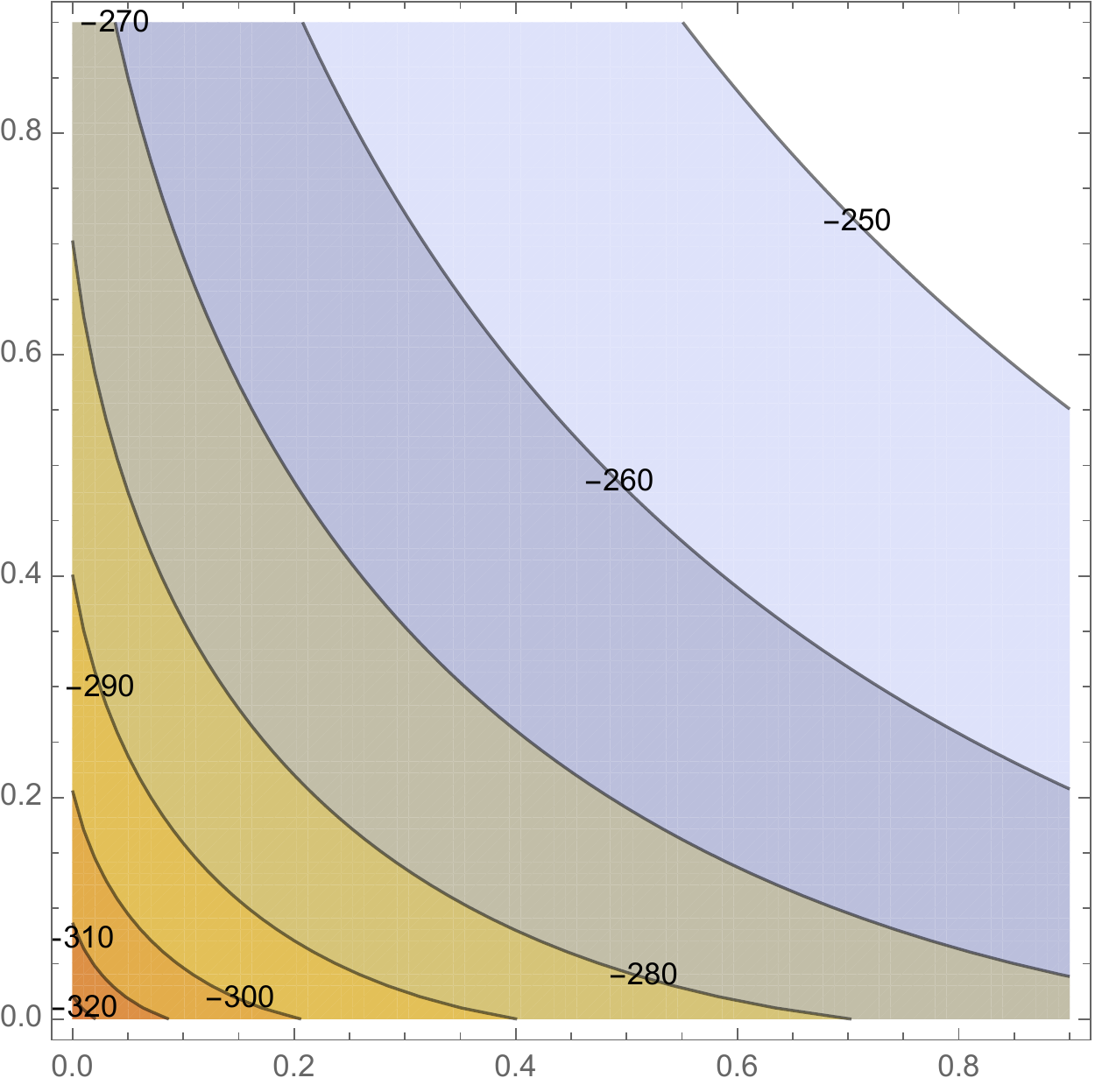} & \includegraphics[scale=0.4]{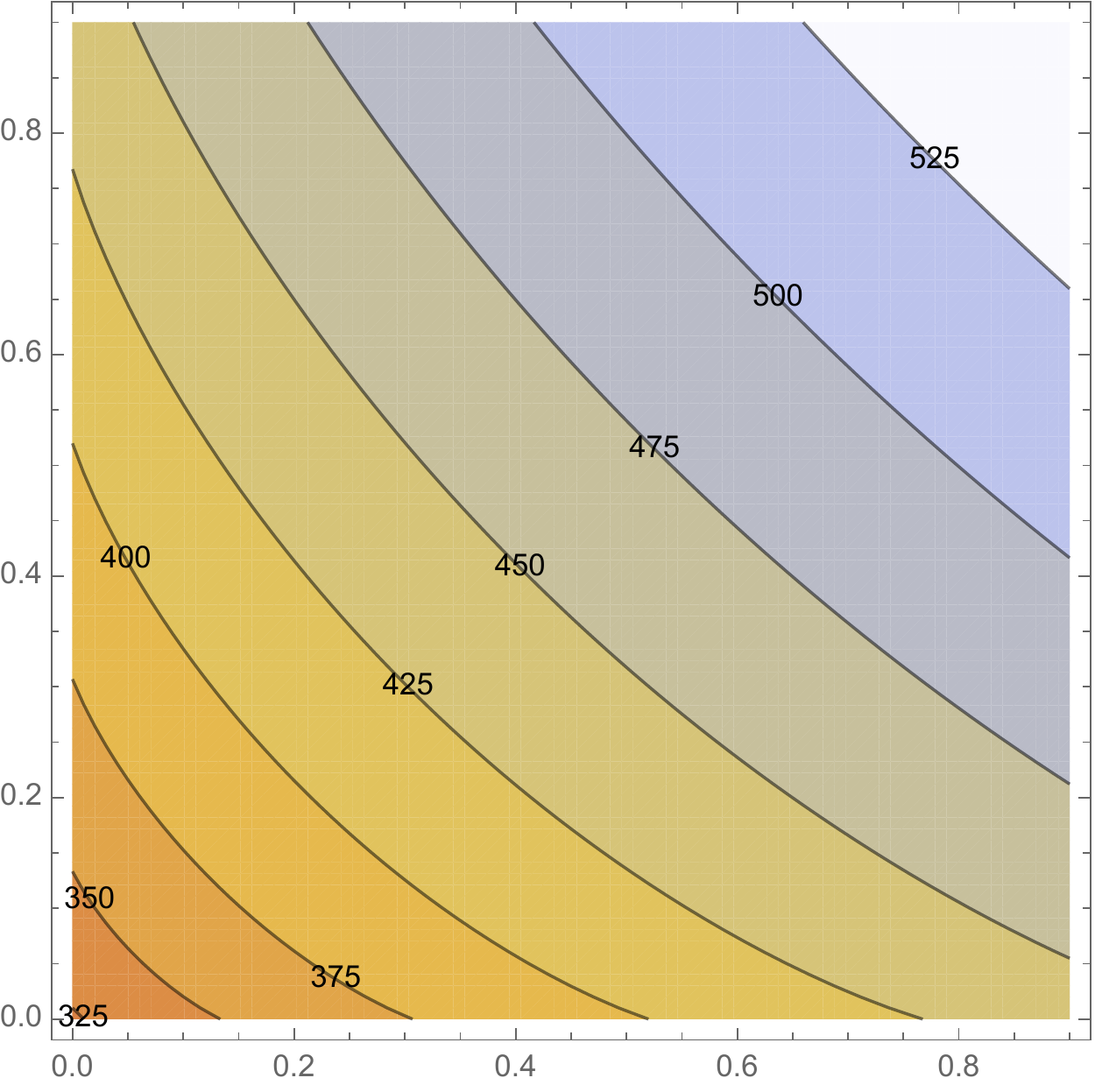}  \\
$d=3.8$ & $d=3.9$ \\
\includegraphics[scale=0.4]{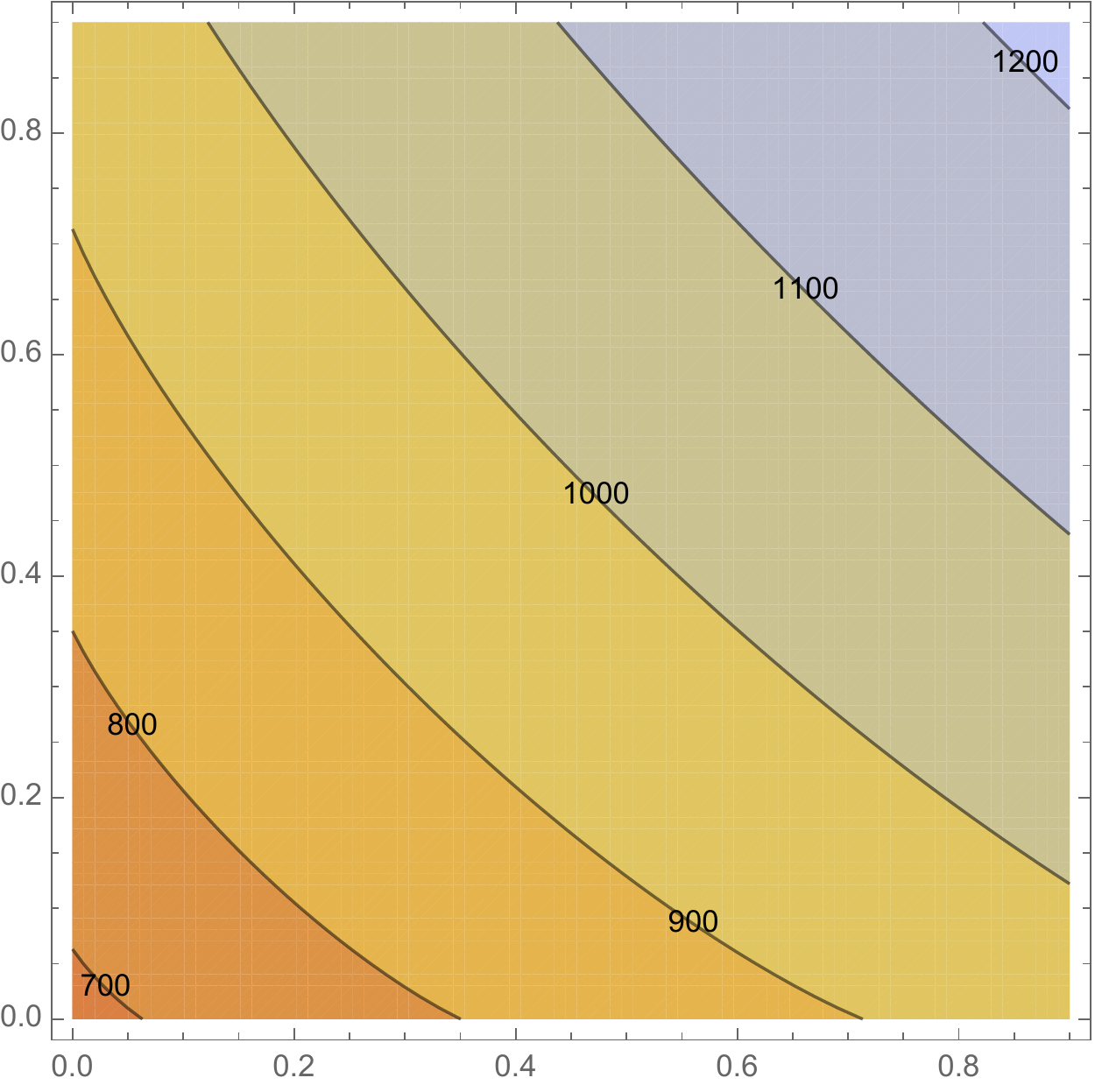} & \includegraphics[scale=0.4]{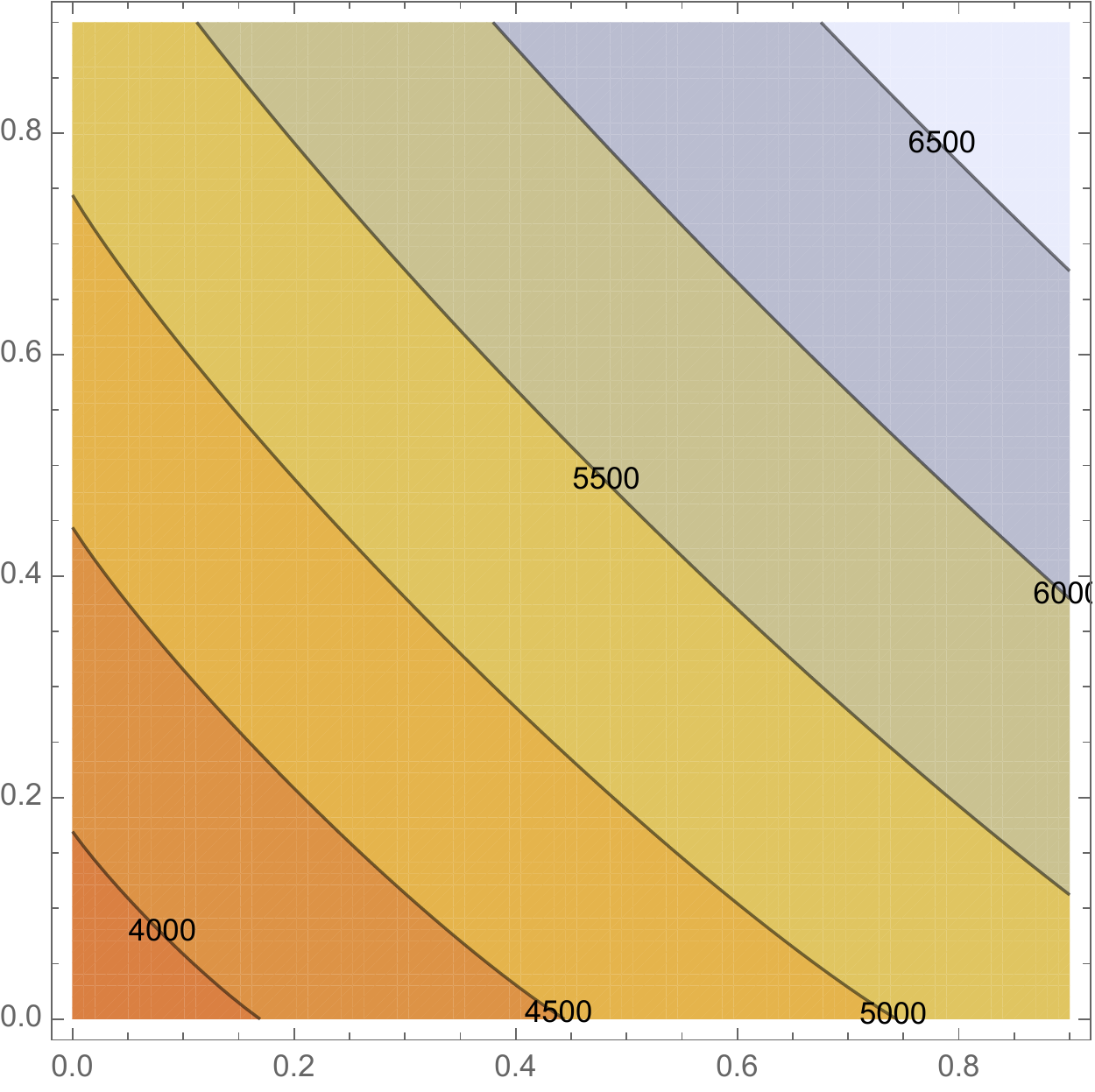}  \\
$d=4.1$ & $d=4.2$ \\
\includegraphics[scale=0.4]{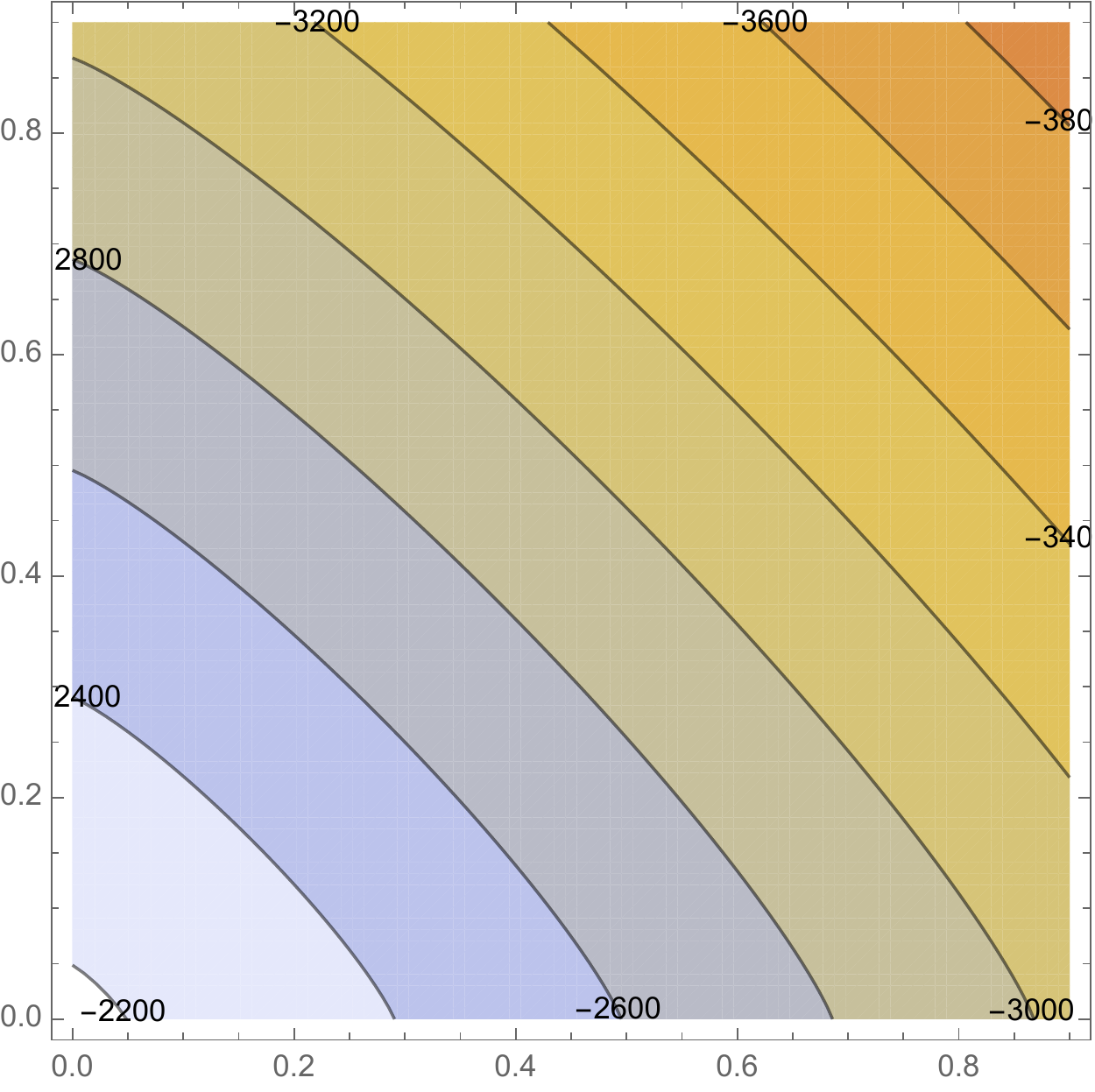} & \includegraphics[scale=0.4]{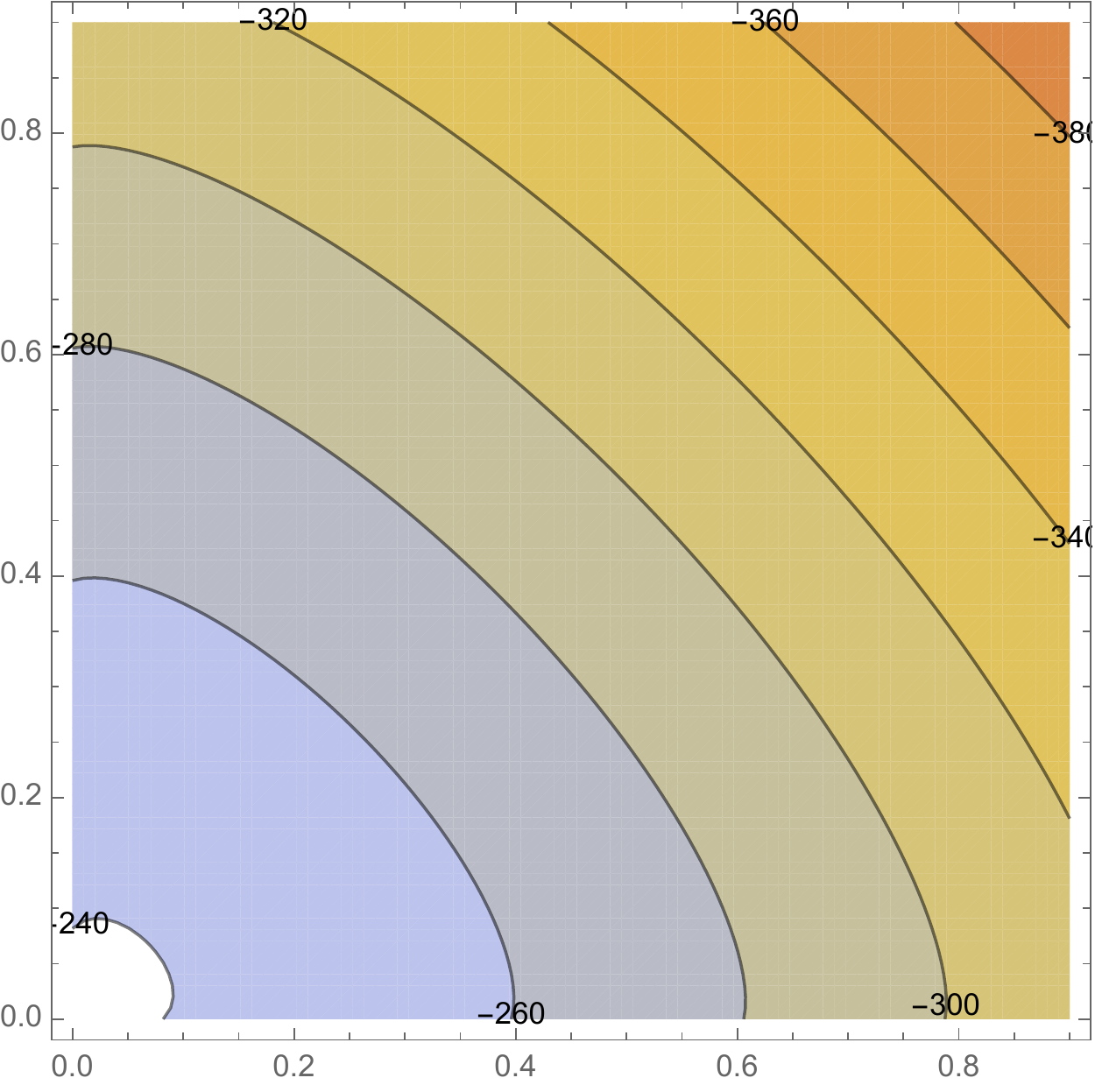}  
\end{tabular}
\end{center}
\label{numerical}
\end{table}

\section{Summary}
\label{sec:summary}
In this section we summarize our results \begin{itemize}

\item We applied the SFI method to the vacuum seagull. The group $G$ is given by (\ref{generators1}) and equivalently by the defining relations (\ref{comrel}); the $G$ foliation is found to be co-dimension 0 (\ref{codim}). The SFI differential equation set is given in (\ref{PDEs}). Finally, the expression for the vacuum seagull diagram in terms of a line integral over simpler diagrams is given in (\ref{cb}-\ref{gensolb}). A concrete 1-dimensional realization for the line integral is given in (\ref{mainlineint}).

\item We presented a geometrical method to determine $G$ as the stabilizer of the space of squares inside the space of quadratics, see Section \ref{sec:eq} as well as \cite{Kol:2016veg}. We presented the first example for this method in the presence of potential numerators (the vacuum seagull has one potential numerator). This method provides the geometry underlying the definition of the group $G$  introduced in \cite{Kol:2016hak}, and allows for a simplified determination of $G$.

\item We obtained explicit results for an integral with 3 mass scales (in the reduced parameter space) in terms of a (rare) special function (\ref{vs_3m}), or alternatively in terms of an explicit 1 dimensional integral (\ref{explicit}). This might be the first explicit  determination of a 3-loop diagram with 3 arbitrary mass scales. We also supplied numerical results over the reduced parameter space (see Table \ref{numerical}) and some terms in the $\epsilon$-expansion (\ref{e-3}, \ref{e-2}).

\end{itemize}

This work adds to that of the closely related \cite{Martin:2016bgz} not only by the last item above, but also by presenting the equation set explicitly, as well as by determining the group structure and the $G$-foliation of parameter space, thereby exposing the underlying geometry. 

\subsection*{Acknowledgments}
We are grateful to A. Davydychev, A. Freitas, S. Martin, D. Robertson and V. Smirnov for useful comments on a draft. We also thank M. Kalmykov and A. Kotikov for useful correspondence.
This research was supported by the ``Quantum Universe'' I-CORE program of the Israel Planning and Budgeting Committee.

\appendix
\section{Appendix}
\subsection*{Source integrals}
$J_{ij}$ are the integrals obtained from equation (\ref{original_integral}) by contracting the propagator $i$ and squaring the propagator $j$.
\begin{eqnarray} 
J_{12}&=&\int \frac{d^d l_1 d^d l_2 d^d l_3}{(l_2^2-x_2)^2((l_1+l_2)^2-x_3)(l_3^2-x_4)((l_1+l_3)^2-x_5)} \label{j12}\\
J_{13}&=&\int \frac{d^d l_1 d^d l_2 d^d l_3}{(l_2^2-x_2)((l_1+l_2)^2-x_3)^2(l_3^2-x_4)((l_1+l_3)^2-x_5)} \label{j13}\\
J_{14}&=&\int \frac{d^d l_1 d^d l_2 d^d l_3}{(l_2^2-x_2)((l_1+l_2)^2-x_3)(l_3^2-x_4)^2((l_1+l_3)^2-x_5)} \label{j14}\\
J_{15}&=&\int \frac{d^d l_1 d^d l_2 d^d l_3}{(l_2^2-x_2)((l_1+l_2)^2-x_3)(l_3^2-x_4)((l_1+l_3)^2-x_5)^2}\label{j15}\\
J_{23}&=&\int \frac{d^d l_1 d^d l_2 d^d l_3}{(l_1^2-x_1)((l_1+l_2)^2-x_3)^2(l_3^2-x_4)((l_1+l_3)^2-x_5)}\label{j23}\\
J_{32}&=&\int \frac{d^d l_1 d^d l_2 d^d l_3}{(l_1^2-x_1)(l_2^2-x_2)^2(l_3^2-x_4)((l_1+l_3)^2-x_5)}\label{j32}\\
J_{45}&=&\int \frac{d^d l_1 d^d l_2 d^d l_3}{(l_1^2-x_1)(l_2^2-x_2)((l_1+l_2)^2-x_3)((l_1+l_3)^2-x_5)^2}\label{j45}\\
J_{54}&=&\int \frac{d^d l_1 d^d l_2 d^d l_3}{(l_1^2-x_1)(l_2^2-x_2)((l_1+l_2)^2-x_3)(l_3^2-x_4)^2}~. \label{j54}
\end{eqnarray}
The integrals (\ref{j12}-\ref{j15}) are of the ``watermelon" topology discussed in the introduction; these integrals and their associated ``sunset" diagrams can be represented in terms of 1-dimensional integrals of Bessel functions \cite{Groote:2005ay, Berends:1993ee, Mendels:1978wc}.
The integrals (\ref{j23}-\ref{j54}) have the topology of a product of the 2-loop ``diameter" diagram and a tadpole. 

\subsection*{Expressions for sources on reduced parameter space} \label{appendix}
On the 3 dimensional subspace discussed in Section \ref{sec:3d} all of the sources can be obtained from the following vacuum two loop integral (diameter topology)
\begin{equation}
J(n_1,n_2,n_3;x_1,x_3;d)=\int \frac{d^dl_1 d^d l_2}{(l_1^2-x_1)^{n_1}((l_2)^2)^{n_2}((l_1+l_2)^2-x_3)^{n_3}}~.
\end{equation}
Equation (4.3) in \cite{Davydychev:1992mt}  gives the result for this two loop vacuum integral with three different masses. By taking one of the masses to zero we get the following expressions
\begin{eqnarray}
& & J(n_1,n_2,n_3;x_1,x_3;d)= \pi^d i^{2 - 2 d} (-x_3)^{
  d - n_1 - n_2 - n_3} \frac{1}{
   \Gamma(n_1) \Gamma(n_2) \Gamma(n_3) \Gamma(
     d/2)} \\
     & & \Big[\Gamma \big(\frac{d}{2} - n_1\big) \Gamma \big(\frac{d}{2} - n_2\big) \Gamma \big(
      n_1 + n_2 - \frac{d}{2}\big) \Gamma \big(n_1 + n_2 + n_3 - d\big) \nonumber \\
      & & \times {_{2}F_{1}}\big(
      n_1 + n_2 + n_3 - d, n_1 + n_2 - \frac{d}{2}, n_1 - d/2 + 1 \big|
      \frac{x_1}{x_3}\big) \nonumber \\
      & &+ \big(\frac{x_1}{x_3}\big)^{d/2 - n1}
      \Gamma \big(n_1 - \frac{d}{2}\big) \Gamma \big(\frac{d}{2} - n_2\big) \Gamma(n_2) \Gamma \big(
      n_2 + n_3 - \frac{d}{2}\big) {_{2}F_{1}}\big(n_2, n_2 + n_3 - \frac{d}{2}, 
      \frac{d}{2} - n_1 + 1 \big| \frac{x_1}{x_3}\big)\Big]~. \nonumber
\end{eqnarray}

Using this formula we can write the explicit form of $J_{12},J_{14},J_{32}$ and $J_{54}$
\begin{eqnarray}
J_{12}&=&\frac{i}{\pi^{3 d/2} e^{3\gamma_E(4 - d)/2}} \pi^{d/2} i^{
 1 - d} \Big(\frac{\Gamma \big(2 - \frac{d}{2}\big)\Gamma \big(\frac{d}{2} - 1\big)^2}{\Gamma(d - 2)}\Big) J(2, 
  2 - d/2, 1; x_2, x_4;d) \label{redj12}\\
J_{14}&=&\frac{i}{\pi^{3 d/2} e^{3\gamma_E(4 - d)/2}} \pi^{d/2} i^{
 1 - d} \Big(\frac{\Gamma \big(2 - \frac{d}{2}\big)\Gamma \big(\frac{d}{2} - 1\big)^2}{\Gamma(d - 2)}\Big) J(2, 
  2 - d/2, 1; x_4, x_2;d) \label{redj14}\\
J_{32}&=& \frac{i}{\pi^{3 d/2} e^{3\gamma_E(4 - d)/2}} \pi^{d/2} i^{
 1 - d} \Gamma \big(2 - \frac{d}{2}\big) (-x_2)^{d/2 - 2} J(1, 1, 1; x_1, x_4; d) \label{redj32}\\
J_{54}&=& \frac{i}{\pi^{3 d/2} e^{3\gamma_E(4 - d)/2}} \pi^{d/2} i^{
 1 - d} \Gamma \big(2 - \frac{d}{2}\big) (-x_4)^{d/2 - 2} J(1, 1, 1; x_1, x_2; d)~. \label{redj54}     
\end{eqnarray}

\subsection*{Generalized Hypergeometric functions} \label{hypergeometric}
The hypergeometric function is defined by the power series
\be
{_{2}F_{1}}(a,b,c|x)=\sum_{n=0}^{\infty}\frac{(a)_n(b)_n}{(c)_n}\frac{z^n}{n!} \label{2f1}
\ee
where $(a)_n=a(a+1)(a+2)\dots (a+n-1)$ with $(a)_0=1$ is the Pochhammer symbol.

This definition can be generalized to a double series depending on two variables. The Appell function $F_1$ is defined by the double series
\be
F_1(a,b_1,b_2,c|x,y)=\sum_{k=0}^{\infty}\sum_{n=0}^{\infty} \frac{(a)_{k+n}(b_1)_k(b_2)_n}{(c)_{k+n}}\frac{x^k}{k!}\frac{y^n}{n!}~.
\ee

The definition (\ref{2f1}) is also generalizes to the generalized hypergeometric function
\be
{_{A}F_{B}}(a_1,a_2,\dots ,a_A,b_1,b_2,\dots ,b_B|x)=\sum_{n=0}^{\infty}\frac{(a_1)_n(a_2)_n\dots(a_A)_n}{(b_1)_n(b_2)_n\dots (b_B)_n}\frac{z^n}{n!}~. \label{afb}
\ee
The Kamp\'e de F\'eriet function \cite{Kampedeferiet} (see also the appendix of \cite{Boos:1990rg}) is a further generalization of (\ref{afb}) to a double series depending on two variables, defined by
\bea
& &F^{\substack{A:B;B'\\ C:D;D'}}\left( \begin{matrix}a_1,\dots ,a_A:b_1, \dots ,b_B; b'_1, \dots ,b'_{B'} \\ c_1 ,\dots, c_C: d_1, \dots ,d_D ; d'_1 ,\dots ,d'_{D'}\end{matrix} \Big| x,y\right)\nonumber\\ &&=\sum_{k=0}^{\infty}\sum_{n=0}^{\infty}\frac{(a_1)_{k+n}\dots(a_A)_{k+n}(b_1)_k \dots (b_B)_k (b'_1)_n \dots (b'_{B'})_n}{(c_1)_{k+n}\dots(c_C)_{k+n}(d_1)_k \dots (d_D)_k (d'_1)_n \dots (d'_{D'})_n}\frac{x^k}{k!}\frac{y^n}{n!}~. \label{Kampedeferiet}
\eea

\bibliographystyle{unsrt} 

\end{document}